\documentclass[manuscript,acmsmall]{acmart}
\setcopyright{cc}
\setcctype{by}
\acmJournal{TOSEM}
\acmYear{2026} \acmVolume{1} \acmNumber{1} \acmArticle{}
\acmMonth{1} \acmDOI{10.1145/3844507}
\usepackage[utf8]{inputenc}
\usepackage[T1]{fontenc}
\usepackage{array}
\usepackage{booktabs}
\usepackage{multicol}
\usepackage{multirow}
\usepackage{tabularx}
\usepackage{longtable}
\usepackage{threeparttable}
\usepackage{enumitem} 
\usepackage{tcolorbox}
\usepackage[super]{nth}
\tcbuselibrary{breakable}
\usepackage{balance}
\usepackage{hyperref}
\usepackage{tikz}
\usepackage{color,soul}
\usepackage{amsmath}
\usepackage{algorithm}
\usepackage[noend]{algpseudocode}
\usepackage{balance}
\usepackage[toc,page]{appendix}
\usepackage{fixltx2e}

\usepackage{todonotes}
\newcounter{todocounter}

\newcommand{\projName}{\textit{SAT-DLink-Test}}
\newcommand{\projNameNew}{\textit{SAT-DLink}}

\newcommand{\sectopic}[1]{\vspace{0.2em}\par\noindent{\textit{\bfseries #1}}}

\newcommand*\circled[1]{\tikz[baseline=(char.base)]{
            \node[shape=circle,draw,inner sep=0.7pt] (char) {#1};}}
            
\newcommand*\colorcircled[1]{\tikz[baseline=(char.base)]{
            \node[shape=circle,draw,inner sep=0.7pt,fill=gray] (char) {#1};}}

\def\BibTeX{{\rm B\kern-.05em{\sc i\kern-.025em b}\kern-.08em
    T\kern-.1667em\lower.7ex\hbox{E}\kern-.125emX}}
\begin{document}

\title{LLM-Driven Cost-Effective Requirements Change Impact Analysis
}

\author{Romina Etezadi}
\affiliation{\institution{School of Electrical Engineering and Computer Science, University of Ottawa}
 \city{Ottawa}
 \country{Canada}}
\email{retez068@uottawa.ca}

\author{Sallam Abualhaija}
\affiliation{\institution{SnT Centre for Security, Reliability, and Trust, University of Luxembourg}
  \city{Luxembourg}
  \country{Luxembourg}}
\email{sallam.abualhaija@uni.lu}

\author{Chetan Arora}
\affiliation{\institution{Faculty of Information Technology, Monash University}
  \city{Melbourne}
  \country{Australia}}
\email{chetan.arora@monash.edu}

\author{Lionel Briand}
\affiliation{\institution{School of Electrical Engineering and Computer Science, University of Ottawa}
  \city{Ottawa}
  \country{Canada}
  }
  \affiliation{\institution{Lero Centre, University of Limerick}
  \city{Limerick}
  \country{Ireland}
  }
\email{lbriand@uottawa.ca}

\begin{abstract}
Requirements are inherently subject to change throughout the software development lifecycle. Within the limited budget available to requirements engineers, manually identifying the impact of such changes on other requirements is error-prone and effort-intensive, especially in regulated domains. This can lead to overlooked impacted requirements, which, if not properly managed, can cause serious issues in downstream tasks.
Inspired by the growing potential of large language models (LLMs) across diverse domains, we propose \textit{ProReFiCIA}, an LLM-driven approach to automatically identify impacted requirements when changes occur. We conduct an extensive evaluation of \textit{ProReFiCIA} using several LLMs and prompt variants tailored to this task. Using the best LLM-prompt combination, \textit{ProReFiCIA} achieves 85.7\% recall on an unseen industrial dataset, demonstrating its effectiveness in identifying impacted requirements. Further, the cost of applying \textit{ProReFiCIA} remains small, as the engineer only needs to review the predicted impacted requirements, which represent 3.0\% of the entire set of requirements. Lastly, incorporating domain knowledge via RAG increases recall to 95.8\% while slightly raising the cost to 3.4\%.

\end{abstract}


\maketitle

\section{Introduction}~\label{sec:introduction}

Software systems today are increasingly complex, distributed, and tightly coupled, making even seemingly minor changes potentially risky and costly. In this context, Change Impact Analysis (CIA) plays a critical role in Requirements Engineering (RE) by systematically identifying, evaluating, and managing the effects of proposed modifications~\cite{jonsson2005impact}. The primary objective of CIA is to determine which components—ranging from requirements and design artifacts to source code and documentation—are likely to be affected by a change in requirements. Accurate impact assessment is essential not only for minimizing unintended side effects but also for supporting cost estimation, prioritization, and strategic decision-making in software maintenance and evolution~\cite{Arora:15b}. 
Throughout the software development lifecycle, requirements often change rapidly due to evolving project needs, stakeholder input, or external factors. When a requirement is modified, it is thus essential to assess the potential impact of this change on other related requirements. This process is particularly crucial in regulated domains (e.g., automotive, medical instruments, energy) and is known as inter-requirement change impact analysis~\cite{pohl2016requirements}. Identifying, assessing, and incorporating these changes across the entire requirements specification can be time-consuming and costly. 
 
Traditionally, CIA has relied on manual inspections, expert judgment, and syntactic dependency analysis. While effective in small systems, these approaches often struggle to scale to large, evolving software landscapes and may miss implicit or semantic dependencies.
To address the challenges of manually identifying impacted requirements, researchers have proposed automated approaches~\cite {nejati2016automated,abualhaija2024toward}. However, research on impact analysis of inter-requirement changes remains limited~\cite{Arora:15Tool}. 
This problem is particularly critical in settings where requirements serve as a contractual basis between stakeholders or where systems are safety-critical and must strictly comply with regulatory standards~\cite{sommerville2011software,wiegers2013software}. In such environments, even minor changes to a single requirement can propagate across dependent requirements, potentially affecting system behavior, compliance, or safety guarantees. As a result, understanding and predicting inter-requirement dependencies is essential.
Other studies on traceability link recovery (TLR), which focus on predicting trace links between different software artifacts, have proposed techniques that can also be applied to CIA tasks through the use of requirement traceability~\cite{hey2025requirements,vogelsang2025impact,mucha2024systematic,li2023applications,abbas2023relationship}. These approaches leverage semantic similarity, dependency graphs, and machine learning to infer relationships between artifacts.
Despite their utility, general TLR techniques alone cannot fully address the primary challenges of CIA.
One key limitation lies in the nature and variability of change requests. In practice, a change may be triggered either by an updated requirement that modifies existing functionality or by a descriptive explanation, often expressed in natural language (NL), outlining how the change should be implemented across related artifacts~\cite{Arora:15b}. 
General TLR techniques, which operate at the requirements level, typically recover relationships or links between requirements using the same NL style rather than focusing on complex change descriptions. 
Consequently, they may fail to capture implicit dependencies, indirect impacts, or nuanced semantic relationships that are critical for accurately predicting change propagation.

\begin{figure*}[t]
\includegraphics[width=\textwidth]{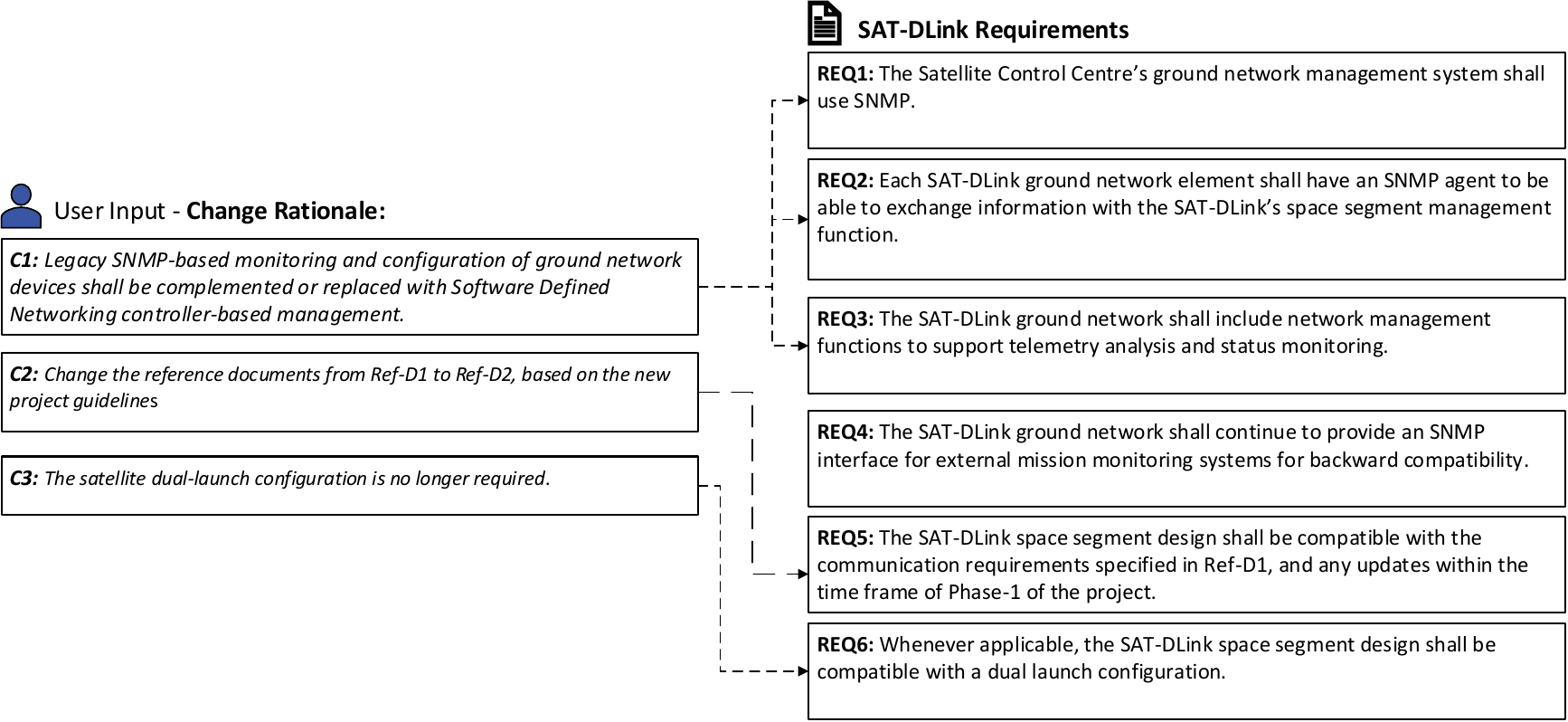}
\caption{Example of three change rationales and their impact on requirements. $C_i$ denotes a change request submitted by the user, and $REQ_i$ represents the $i^{th}$ requirement in the documentation.}
\label{fig:example}
\end{figure*}

To illustrate this concept, consider the example in Figure~\ref{fig:example}. The requirements pertain to the functionality and operations of a satellite data link management system (referred to as \projName, in this paper). As we discuss later in the paper, these requirements are drawn from a real satellite industry requirements document and have been marginally altered to preserve confidentiality.
The user has three distinct change requests: C1, C2, and C3. In C1, the user intends to modify the \projName~program’s functionality by replacing or augmenting the legacy networking protocol SNMP (Simple Network Management Protocol) with SDN (Software-Defined Networking) technology. Following this request, requirements REQ1 and REQ2 are impacted as they directly refer to the use of SNMP in the internal operations of the system. REQ3 is also impacted, as it refers to network management functions affected by SDN technology. REQ4, on the other hand, is not impacted, as its intent is system interoperability rather than internal management, despite discussing SNMP. In C2, the user requests that all documents referencing Ref-D1 (as defined in the requirements document) be updated to a newer version (Ref-D2) aligned with the new project guidelines. Consequently, any requirements that mention Ref-D1 documents, such as REQ5, are impacted by this change. In C3, the user requests removing the dual-launch configuration. Therefore, all requirements associated with launch operations that depend on or reference the dual-launch setup, such as REQ6, will be affected.

The advent of LLM models has opened new avenues for RE tasks~\cite{zadenoori2025large}. LLMs can capture subtle semantic relationships between NL texts and, more importantly, infer indirect dependencies, which is essential in CIA tasks. Building on LLM capabilities and prompt engineering, we introduce \textit{ProReFiCIA}, an approach that leverages LLMs' reasoning and contextual understanding to identify potential change propagation. As a result, \textit{ProReFiCIA} improves the completeness of detected impacts while maintaining high accuracy in predicting which requirements are truly affected.

\sectopic{Contributions.}  
This paper makes the following contributions:
\begin{itemize}
    \item  We design 64 distinct prompt variations and conduct an extensive evaluation to analyze how prompt detail and structure influence LLM performance on the CIA task across datasets spanning different domains and complexity levels.
    \item  We conduct a comprehensive evaluation across five distinct LLMs to assess their relative effectiveness at addressing the CIA task.
    \item We present \textit{ProReFiCIA}, an automated approach for CIA that combines refinement and filtering stages with advanced prompt engineering techniques to improve performance and compare it with baseline techniques. We designed \textit{ProReFiCIA} to minimize the amount of labeled training data required.
\item We evaluate \textit{ProReFiCIA}. On a new, unseen, and more complex industry dataset, we show that, on average, it successfully identifies 85.7\% of the impacted requirements while retrieving only 3.0\% of the total requirements for the analyst's inspection. Results therefore show the high cost-effectiveness of \textit{ProReFiCIA}.
    \item We demonstrate that incorporating domain knowledge through RAG further enhances \textit{ProReFiCIA}, increasing its effectiveness to 95.8\% while requiring the analyst to inspect only 3.4\% of the total requirements.
\end{itemize}

The structure of the paper is as follows: Section~\ref{sec:background} provides background information. Section~\ref{sec:approach} presents our proposed approach. Section~\ref{sec:evaluation} reports on our empirical evaluation. Section~\ref{sec:threats} discusses threats to validity. Section~\ref{sec:discussion} discusses the practical implications of our results. Section~\ref{sec:related} reviews the related work, and finally, Section~\ref{sec:conclusion} concludes the paper.

 \section{Background}~\label{sec:background}
In this section, we provide definitions of the key concepts and terms used throughout this paper. The definitions are organized into two categories: 1) Change Impact Analysis (CIA) and 2) Natural Language Processing (NLP).

\subsection{Change Impact Analysis} 
\sectopic{Change Rationale.}\label{change_rationale}
A change rationale is an NL description that specifies a change to be applied to the requirements documents. Furthermore, a change rationale may include a modified requirement~\cite {Arora2015}. The description of a change rationale can belong to three categories~\cite{Arora2015}:
1)~\textbf{Addition:} In this category, the description of a change indicates the addition of a new feature or functionality. For example, consider the \projName~system introduced in Section~\ref{sec:introduction}, with the following change rationale: ``\textit{The \projName~platform shall support another ancillary payload (AP-1) up to 2.2 kg in weight.}''. This statement indicates that a new payload (AP-1) is to be integrated into the system. Consequently, adding this payload may affect other requirements related to the platform’s payload capacity, system interfaces, operational procedures, and resource allocations.
2)~\textbf{Deletion:} The description indicates that a specific feature or functionality is to be removed from the system. For example, the change rationale ``\textit{dual launch configuration is not required}'' indicates that the dual launch capability should be eliminated. 
3) \textbf{Modification:} This category involves both the removal of an existing feature or functionality and the addition of a new one. For example, consider the change rationale: ``\textit{Change the reference documents from Ref-D1 to Ref-D2, based on the new project guidelines}''. In this case, any previous reference to Ref-D1 should be replaced with the new project reference document, i.e., Ref-D2. Such modifications may have cascading effects on related system components.
\sectopic{Impact Set.}
When a change rationale is provided, it is necessary to identify and list all requirements that may be affected by the proposed change. The collection of such requirements is referred to as the \textit{impact set}. This impact set represents the full scope of potential changes within the system, capturing both direct and indirect effects on the requirements. Correctly determining the impact set is essential to maintain system consistency, ensure traceability, and support informed decision-making during the change management process.

\subsection{Natural Language Processing}
\sectopic{Large Language Models.} LLMs have revolutionized NLP by leveraging the Transformer~\cite{Vaswani:17} architecture, which employs self-attention mechanisms to capture intricate dependencies within data sequences. Subsequent advances have led to LLMs with tens to hundreds of billions of parameters, trained on vast corpora using self-supervised learning techniques~\cite{wu2024continual}. Therefore, instead of fine-tuning language models (LMs), prompt engineering reformulates downstream tasks as natural-language instructions.
With the right prompts, a single unsupervised LLM can handle many tasks without additional training. However, the model's effectiveness depends on carefully designed input prompts~\cite{liu2023pre,wu2024continual}.
\sectopic{LLM Embeddings.}
Embeddings derived from LLMs (LLM Embeddings) represent a new generation of contextual representations that leverage the scale and multi-task capabilities of modern LLMs. Unlike embeddings from traditional pretrained language models (PLMs) such as BERT, RoBERTa, or ALBERT, LLM embeddings come from models trained on massive corpora with autoregressive or instruction-following objectives. This broader training enables LLM embeddings to capture not only syntactic and semantic properties but also higher-level reasoning, world knowledge, and contextual subtleties~\cite{krassovitskiy2025llm}, making them better suited to tasks that require nuanced semantic understanding, complex instructions, or domain-specific knowledge.
\sectopic{Retrieval-Augmented Generation.} LLMs have achieved significant success across many NLP tasks; however, they still face limitations in scenarios that require domain-specific knowledge~\cite{kandpal2023large}. To address this challenge, Retrieval-Augmented Generation (RAG)~\cite{lewis2020retrieval,guu2020retrieval} has been proposed to incorporate domain knowledge into the model. With recent advances in prompt engineering, RAG retrieves relevant information for a query and incorporates it into the prompt as additional context. This allows the model to access external, domain-specific knowledge beyond its training data, leading to more accurate, grounded, and context-aware responses. 
\sectopic{Cache-Augmented Generation.} Recent advances in LLMs capable of processing longer texts have prompted a shift toward cache-augmented generation (CAG)~\cite{chan2025don}. This approach addresses the traditional retrieval bottleneck by preloading relevant knowledge directly into the LLM’s extended context window, enabling more efficient access to necessary information~\cite{huang2023towards}. By incorporating information within the context window, CAG can also improve response coherence, ensuring each part of the generated output stays aligned with the provided context rather than drifting into unrelated content~\cite{mialon2023augmented}. Despite these benefits, processing longer text sequences introduces challenges, such as the "lost-in-the-middle" problem, where the model may inadvertently overlook or underweigh portions of the prompt~\cite{liu2024lost}.
\sectopic{Natural Language Inference.} Natural Language Inference (NLI)~\cite{dagan2005pascal,jm3} is a fundamental task in natural language processing (NLP) that involves determining the logical relationship between a premise and a hypothesis. The goal is to classify the relationship into three categories: entailment, contradiction, or neutral. For instance, consider the premise ``\textit{A man is eating a salad}''. From this statement, we can infer the hypothesis, ``\textit{A man is having a meal}''. In this case, the premise entails the hypothesis, since eating a salad constitutes a meal. This illustrates how NLI involves reasoning about the logical relationship between two texts. \section{Approach}~\label{sec:approach}
\begin{figure*}
\includegraphics[width=0.8\textwidth]{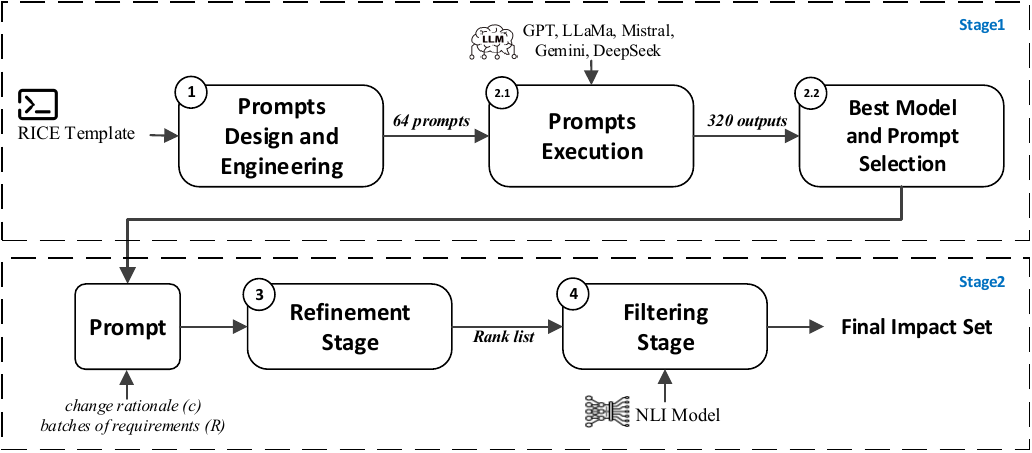}
  \centering
\caption{Overview of the steps in the study. Stage 1 identifies the most effective and generalizable combination of model and prompt. Stage 2 presents our proposed approach, \textit{ProReFiCIA}, which automatically generates an impact set for a given change rationale.}
\label{fig:cia-approach}
\end{figure*}

In this section, we describe the key components of \textit{\textbf{ProReFiCIA}} (Prompt-Refinement-Filtering for CIA). Figure~\ref{fig:cia-approach} depicts our two-stage approach. As shown in the figure, Stage 1 (steps 1 and 2) covers the preparatory steps carried out to optimize \textit{\textbf{ProReFiCIA}}, and Stage 2 (steps 3 and 4) focuses on refinement and filtering.
In Step 1~\ref{sec:prompt_engineering}, we show how prompt engineering is performed, with an emphasis on optimizing model outputs to ensure relevance, accuracy, and alignment with the intended objectives. 
In Step~2~\ref{sec:execution}, we provide a brief overview of how the prompts are applied to the LLM models.
In Step 3~\ref{sec:refinement} and 4~\ref{sec:filtering}, we introduce our proposed post-processing methods, which consist of two steps, Refinement and Filtering, applied to finalize the impact set.

\begin{figure*}[t]
\includegraphics[width=\textwidth]{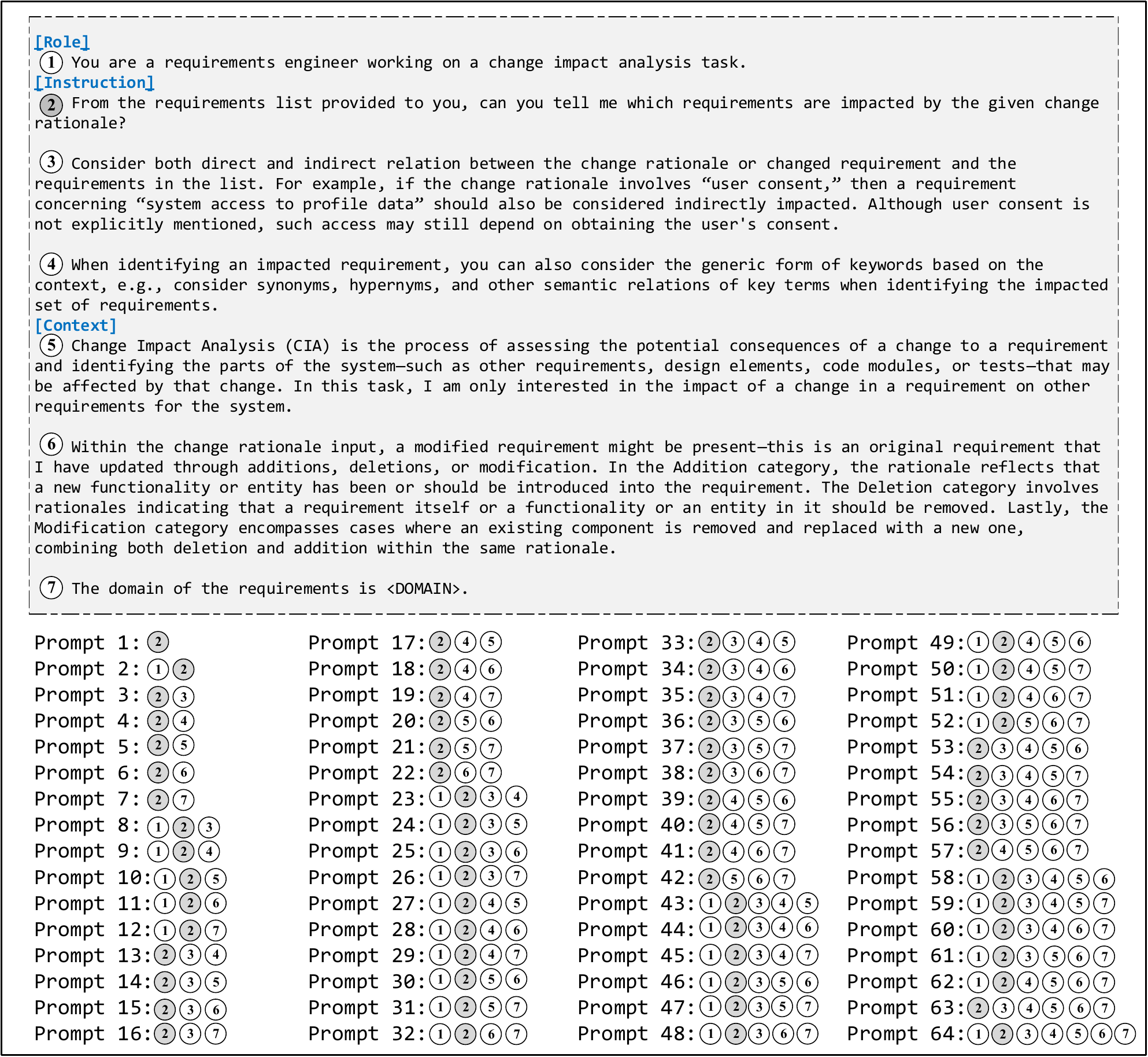}
\caption{Overview of the development of the prompts. Text 2 (gray circle) in the instruction section is mandatory for the prompt to define the task.}
\label{fig:prompts-flow}
\end{figure*}

\subsubsection{\textbf{Step 1: Prompt Engineering}}\label{sec:prompt_engineering}
To identify optimal prompts, we systematically explore different prompt variations using the RICE template~\cite{vogelsang2024using}. The RICE template is selected to incorporate the most commonly used prompt components, namely role, instruction, and context, in our study, which are often considered foundational for guiding LLM behaviour~\cite{liu2023pre,sahoo2024systematic,huang2025prompt}. The role defines the model’s behavior and perspective, the instruction clearly specifies the task, and the context provides necessary background information. Together, these components form a foundational structure that improves control, clarity, and relevance in LLM outputs.
We selectively include or exclude different elements depending on the desired level of detail. The RICE template consists of four components: Role, Instruction (i.e., \textit{What the model is told to do}), Context or Constraints (i.e., \textit{The setting and rules for how to do it}), and Examples for a few-shot setting. In this paper, we exclude examples to evaluate LLMs in a purely zero-shot setting, ensuring that performance reflects their inherent generalization abilities without reliance on task-specific data.
Each component of the RICE template can include varying levels of detail, which can influence the LLM's outcome. 
Accordingly, we design task-relevant details to align with the CIA task's objectives for each prompt component.
\sectopic{\textbf{Role:}} In the context of prompt engineering, \textit{role} typically refers to the persona or perspective the model is instructed to take on during its response. However, this component is optional in our approach, and we must assess its effectiveness as part of the prompt. The text labeled Detail~\circled{1} in Figure~\ref{fig:prompts-flow} illustrates the \textit{role} component by asking the model to adopt a requirements-engineer persona.
\sectopic{\textbf{Instruction:}} In prompt engineering, \textit{instruction} refers to the explicit directive given to the model, telling it what to do. For the CIA task, we propose three distinct directives. Among them, the directive corresponding to the core CIA objective (Detail~\colorcircled{2}) is considered mandatory and is assumed to be always included in the prompt. Details~\circled{3} and~\circled{4} correspond to two optional instructions included in the prompt. Detail~\circled{3} is intended to encourage the model to consider commonsense knowledge that may be necessary to understand the impact of the change. Detail \circled{3} captures key CIA concepts in a single instruction. By providing the example linking "user consent" to "system access to profile data," it emphasizes semantic dependencies, where relationships rely on common sense rather than explicit wording, which is particularly important when change rationales use different terms than the requirements. Finally, encouraging the model to consider such indirect links implicitly supports impact propagation, extending reasoning beyond requirements that are immediately changed.
Detail~\circled{4}, on the other hand, guides the model towards identifying semantic relationships between keywords that appear in both the change rationale and the requirements. Instructing the model to consider synonyms, hypernyms, and other semantic relations enables it to identify requirements that are conceptually related but may not share the exact wording of the change rationale. This addresses situations where impact cannot be detected through direct textual matches.
\sectopic{\textbf{Context:}} In prompt engineering, \textit{context} refers to the background information or supporting content provided to the model. It helps the model better understand the situation, task, or domain, leading to more accurate and relevant responses. For this component, we define three primary types of context. First, Detail~\circled{5} provides a brief explanation of the CIA task. This detail helps the model better understand the CIA task by providing a high-level description of what change impact analysis entails and clarifying the task scope, which is inter-requirement CIA.
Second, Detail~\circled{6} offers a description of what a change rationale is, including the types of information it may contain. This detail provides a structural characterization of change rationales, explaining how a requirement may be modified through addition, deletion, or modification. This helps the model correctly interpret the change rationale before performing impact analysis.
Third, Detail~\circled{7} specifies the domain in which the requirements are written. This detail helps the model to use its internal knowledge to better interpret domain-specific terms and concepts used in both the change rationale and the requirements.
All details within the \textit{context} component are optional. We generate all possible prompt variations by systematically including or excluding each optional detail.
Figure~\ref{fig:prompts-flow} lists the 64 prompts resulting from this process. For example, prompt 12 only includes details~\circled{1} and \circled{7} along with the main instruction~\colorcircled{2}.

\subsubsection{\textbf{Step 2: Prompt Execution and Selection}} \label{sec:execution}
With 64 prompt variations and 5 LLM models, we apply the CAG technique described in Section~\ref{sec:background}, providing each model with the full list of requirements. The model can process all requirements, thereby reducing the number of queries. 
The robustness of LLM models and the effect of each prompt detail are examined in RQ1 and RQ2, respectively, as discussed in Sections~\ref {sec:RQ1} and~\ref {sec:RQ2}.
In addition, given the large number of prompt variations and LLM models, and the high cost of evaluating all combinations, we select the model–prompt pairs for investigation in post-processing. This selection process is the focus of RQ3, elaborated in Section~\ref{sec:RQ3}.

\subsubsection{\textbf{Step 3: Refinement Step}} \label{sec:refinement}
In the refinement step, we address the tendency of LLM models to occasionally overlook certain requirements in the list. Therefore, motivated by the way humans iteratively revisit overlooked information to ensure completeness, we introduce an additional post-processing step to mitigate the lost-in-the-middle problem discussed in Section~\ref{sec:background}. This step reuses the same prompt template, applying it to a shortened list of requirements that were not selected in the initial LLM execution, providing another opportunity for consideration. 
This step is called the refinement step. The requirements identified in this step are then combined with those from the first execution to form the updated impact set. Below, we present an example illustrating how this step is applied:

\begin{tcolorbox}[arc=1mm,width=\columnwidth,top=0mm,left=0mm,  right=0mm, bottom=0mm,
                  boxrule=0.3pt,
               colback=gray!15!white,
               colframe=black, breakable, 
]

Using Prompt 15 from Figure~\ref{fig:prompts-flow}:
\texttt{
\newline
\textbf{Input:}
\newline
\colorcircled{2} \circled{3} \circled{6}
\newline
Requirements List:
\newline
R1: <text>, R2: <text>, R3: <text>, R4: <text>, R5: <text>, R6: <text>
\newline
The output should be in the format: impacted ReqID: <ID> justification: <text>
\newline
\textbf{LLM Output:} 
\newline
impacted ReqID: R2,justification: <text>
\newline
impacted ReqID: R3,justification: <text>
\newline
impacted ReqID: R5,justification: <text>
\newline
\newline
\textcolor{black}{\textbf{\textit{Refinement Step:}}}
\newline
\textbf{Input:}
\newline
\colorcircled{2} \circled{3} \circled{6}
\newline
Requirements List:
\newline
R1: <text>, R4: <text>, R6: <text> 
\newline
The output should be in the format: impacted ReqID: <ID> justification: <text>
\newline
Only select those requirements that you are completely certain about.
\newline
\textbf{LLM Output:} 
\newline
impacted ReqID: R1,justification: <text>
\newline
impacted ReqID: R6,justification: <text>
\newline
\newline
\textbf{Final Output:} \newline
impacted ReqID: R1,justification: <text>
\newline
impacted ReqID: R2,justification: <text>
\newline
impacted ReqID: R3,justification: <text>
\newline
impacted ReqID: R5,justification: <text>
\newline
impacted ReqID: R6,justification: <text>
}
\end{tcolorbox}

\subsubsection{\textbf{Step 4: Filtering Step}} \label{sec:filtering}
Following the refinement step, we obtain the impact set corresponding to the change rationale, along with justifications for selecting each requirement. At this step, we can present the generated impact set to a requirements analyst for review.
We further propose the following algorithm to reduce the false positives in the impact set:
\sectopic{Ranking Phase:} In this phase, the LLM is prompted to rank the requirements within the impact set according to its confidence in their impact. To support this, the model receives the change rationale, the full text of each requirement, and the LLM-generated reasoning used to select them. Presenting this contextual information encourages reflection on prior decisions and supports a critical reassessment of outputs. This process enables the LLM to reorder the impact set based on a more thorough evaluation of relevance and confidence, ultimately enhancing the prioritization of requirements for further analysis. Below, we present the prompt we utilize to get the rank list:
\begin{tcolorbox}[arc=1mm,width=\columnwidth,top=0mm,left=0mm,  right=0mm, bottom=0mm,
                  boxrule=0.3pt,
               colback=gray!15!white,
               colframe=black, breakable, 
]

Prompt used to rank the requirements in the impact set:
\texttt{
\newline
You are an analyst in the field of requirements engineering. I will provide a change rationale, its corresponding impact set, and justification for selection. Rank the requirements in the impact set according to the logical relevance to the change rationale, based on the content of the requirement texts and the provided justifications for their selection. Include all requirement IDs in the sorted list.
\newline
Output format: Sorted\_List: <req\_ids>
\newline
Change Rationale: <change\_rationale\_text>
\newline
Impacted Requirements: <req\_ids>
\newline
Justification: <text>
}

\end{tcolorbox}

Note that the LLM used in this stage is the same model that has been used during the first query and the refinement stage, as this ensures consistency in reasoning across selection, refinement, and ranking, while preserving alignment with the model’s own justifications and avoiding the introduction of additional reasoning patterns or errors that may arise from using a different LLM.
\sectopic{Selection Phase:} This phase is described in Algorithm~\ref{alg:selection}, which filters the ranked requirements. It finalizes the impact set by using a natural language inference model described in Section~\ref{sec:background}. It takes two textual inputs and determines whether the second input logically entails the first. In this context, the NLI model assesses whether a selected requirement, together with its accompanying justification, entails the given change rationale. The output is a binary prediction indicating entailment (labelled as 1) or non-entailment (labelled as 0). As this context requires only two labels (impacted or not impacted), we exclude the third label of the NLI model discussed in Section~\ref{sec:background} during fine-tuning.
The details regarding how this fine-tuning process was applied in our experiments are presented in Section~\ref{sec:RQ3}.
To maintain recall while minimizing false positives, we combine the NLI results with the previously obtained ranked list of requirements. This approach balances retaining true positives with filtering out false positives. More precisely, the following algorithm is applied to determine whether a requirement should remain in the final impact set:
\begin{itemize}
    \item If the size of the impact set is less than 5, no cutoff is applied—the entire set of requirements is retained, and the process terminates at this step, as the set is sufficiently small for rapid human inspection.
    \item If the impact set size exceeds 5, the algorithm proceeds as follows: For each requirement, first, the NLI output is examined. If the prediction is 1, the requirement is retained. If the prediction is 0, the algorithm reconsiders the requirement's position in the sorted list. If the position falls within the top 50\%, the NLI output is ignored, and the requirement is retained, as items in the higher-ranked half are more likely to be relevant. Otherwise, we treat the requirement as a false positive and remove it. 
    We based the decision to retain only the top half on our observation that, in preliminary experiments, the impacted requirements mainly appear in the first half of the sorted list.  
\end{itemize}

Combining the ranking step with NLI models helps preserve recall. The ranking component ensures that highly relevant requirements are retained, even if the NLI model fails to detect them. At the same time, when the impacted set contains, for example, eight instances with only two false positives, the ranking step may select only the top four and exclude other valid instances despite their relatively high rank. In such cases, the NLI model can still detect them, so they together yield more complete results. 

\begin{algorithm}
\caption{}\label{alg:selection}
\begin{algorithmic}[1]
\Statex \textbf{Input:} change rationale $C$ and ranked candidate requirements with their justifications $L_c$
\Statex \textbf{Output:} impact set $I$
\Procedure{Selection Phase}{}
\State let $C$ be a change rationale;
\State let $L_c$ be the list of ranked requirements for $C$;
\State let $M$ be the fine-tuned NLI model;
\State $\textit{ListLen} \gets \text{length of } L_c $;
\If {$\textit{ListLen} \leq 5$} \Return $L_c$ 
\EndIf
\State let $I = []$ be the final impact set;
\For{$i \gets 1$ to $\textit{ListLen}$}
\State {$\textit{label} \gets M (C, L_c[i])$}
\If {$\textit{label} = 1 $}
\State {$ I \gets L_c[i]$};
\ElsIf {$i \leq \textit{ListLen}/2 $}
\State {$ I \gets L_c[i]$};
\EndIf
\EndFor
\State \Return $I$ 
\EndProcedure
\end{algorithmic}
\end{algorithm}

 \section{Empirical Evaluation}~\label{sec:evaluation}
This section presents the evaluation of our approach. We begin with a brief description of the baselines and datasets employed in Section~\ref{baseline} and \ref{datasets}, followed by an introduction to the LLM models selected for this study in Section~\ref{model-selection}. Finally, we discuss the research questions and their corresponding conclusions in Section~\ref{subsec:RQs}.

\subsection{Baselines}\label{baseline}
To strengthen our evaluation, we compare our proposed solutions against several baselines. 
In order to select baselines, we rely on the following exclusion and inclusion criteria:
\begin{itemize}
    \item We do not include techniques that rely on training models using large amounts of labeled data. Such approaches, while they may be effective, are not suitable for our problem context due to the limited training data typically available.
    \item We include only fully automated approaches for generating the list of impacted requirements, without a human in the loop, as this typically requires significant resources that are not typically available.
    \item As our technique is prompt-based, we also include existing prompt-based approaches proposed in the CIA or requirements traceability domains.
\end{itemize}
Based on these criteria, we selected two baselines: 

1) The first baseline is based on a traceability technique presented in the literature that employs RAG. 2) The second baseline is based on a straightforward semantic similarity algorithm that leverages embeddings generated by LLMs, as it is straightforward and does not require training.

\sectopic{CoT.}
As part of our evaluation, we consider the recently proposed requirements traceability technique by Hey et al.~\cite{hey2025requirements}. In their approach, they introduced a RAG method that combines a retrieval step with Chain-of-Thought (CoT) prompting. For each source requirement, the RAG retrieves the top-k target requirements that are most semantically similar. The LLM then receives each source–target requirement pair and, via Chain-of-Thought prompting, determines whether a trace link exists between them. For retrieval, they used OpenAI’s embedding model \texttt{text-embedding-3-large}; for prompting, they tested several LLMs, with GPT-4o showing the best overall performance. Moreover, they have experimented with two different prompt structures: \textbf{KISS} and $CoT$. The KISS prompt is a simple zero-shot prompt that asks the LLM a yes-or-no question about whether the source requirement is related to the target requirement. In contrast, the $CoT$ uses chain-of-thought prompting to determine traceability by providing the source and target artifacts and asking, as in \textbf{KISS}, a simple yes-or-no question about whether a link should exist between them. Based on their experiments, $CoT$ showed better average performance. Therefore, in our study, we focus exclusively on GPT-4o and the $CoT$ prompt for this method.

\sectopic{Semantic Similarity (S).}
With the emergence of powerful embedding models derived from LLMs, the need for extensive training or fine-tuning has decreased significantly. In this work, because no task-specific training or fine-tuning data is available, we use LLM embeddings as a baseline for semantic similarity, providing richer, more informative representations. To select an appropriate model, we use the top-ranked embedding model, \texttt{gemini-embedding-001}, based on the MTEB (Massive Text Embedding Benchmark) leaderboard\footnote{https://huggingface.co/spaces/mteb/leaderboard}. We compute the similarity between each change rationale and each requirement as the cosine similarity between their embeddings. For each change rationale, we generate a list of similarity scores for all requirements. To determine the impact set, we rank requirements in descending order by similarity score (highest to lowest) and apply a cutoff to select the most relevant ones. We employ three different strategies for cutoff point selection to systematically evaluate the effect of the cutoff point choice on the resulting impact sets: 
\begin{itemize}
  \item  \textit{T1:} We use a fixed threshold value, and any requirement with a similarity score exceeding this threshold is included in the impact set. For this study, we use the constant value 0.5 because it is the midpoint of the similarity interval.
  \item  \textit{T2:} This cutoff point selection approach, proposed by~\cite{etezadi2025classification}, is applied on a sorted list of requirements. For each change rationale, cosine similarity is computed between the rationale and each requirement, yielding a similarity score for each requirement. The resulting scores are then used to rank the requirements accordingly. The cutoff is determined by analyzing the differences between consecutive similarity scores in the sorted list, in descending order. The cutoff point is set to the position of the largest drop. For example, given similarity scores 0.85, 0.82, 0.80, 0.78, 0.60, 0.58, 0.57, and 0.40, the differences between consecutive scores would be 0.03, 0.02, 0.02, 0.18, 0.02, 0.01, and 0.17. The largest difference, 0.18, occurs between 0.78 and 0.60, indicating a sharp decline. Therefore, we set the cutoff at 0.78, marking the boundary between requirements likely to be impacted by the change and those that are not.
  \item  \textit{T3:} NARCIA~\cite{Arora:15Tool} proposes a specific technique for cutoff point selection, which we also adopt in this work. Similar to \textit{T2}, a potential impact score is assigned to each requirement, and the resulting list is subsequently sorted in descending order. Then, we determine the cutoff point in the sorted list of requirements using a delta chart to visualize differences between consecutive similarity scores. The cutoff is set at the point beyond which no significant peaks remain, meaning the scores no longer clearly distinguish which requirements are affected. A peak is considered significant if it exceeds one-third of the highest peak. For example, consider the example illustrated in \textit{T2}. The highest peak is 0.18 (between 0.78 and 0.60), so any peak larger than 0.18/3 = 0.06 is considered significant. The second peak at 0.17 (between 0.57 and 0.40) is also significant and is the last. The cutoff is set at the end of the right slope of the last significant peak (0.17 between 0.57 and 0.40), so it meets the requirement with a similarity score of 0.57.
\end{itemize}

\begin{table}
\caption{Datasets Information. 
}
\label{tab:datasets}
\begin{tabularx}{\columnwidth}{@{} p{0.2\columnwidth} @{\hskip 1em} p{0.4\columnwidth} @{\hskip 1em} p{0.2\columnwidth} @{\hskip 1em} *{3}{>{\centering\arraybackslash}X}@{}}
    \toprule
     \textbf{Dataset} & \textbf{Description} & \textbf{Domain} & \textbf{Req} & \textbf{C} & \textbf{P} \\
    \midrule
    \textbf{WASP} & Functionalities and services provided by the mobile service platform & Mobile Service & 72 & 5 & 31\%\\
    \midrule
    \textbf{\projNameNew} &\multirow{2}{*}{Satellite data link management system} & \multirow{2}{*}{Satellite} & \multirow{2}{*}{192} & 5 & 14\% \\
    \textbf{\projName} & & & & 11 & 17\% \\
\bottomrule
\end{tabularx}
\begin{tablenotes}
     \item \textit{\textbf{Req:} Number of requirements, \textbf{C:} Number of change rationales, \textbf{P:} Percentage of the requirements in the list being impacted by any of the change rationales}
     \end{tablenotes}
 \end{table}

\subsection{Datasets}\label{datasets}
The proposed methodology is evaluated using two datasets: a publicly available benchmark dataset~\cite{goknil2014change,goknil2014experimental} based on the WASP system requirements documents~\cite{ebben2002requirements} that was generated under the guidance of an industrial expert, and a newly introduced real-world dataset, \projNameNew~which we split into two subsets, \projNameNew~and \projName~. Details of the datasets are presented in Table~\ref{tab:datasets}.
\sectopic{WASP.}
This dataset is derived from a requirements document for a context-aware mobile service platform. It contains five change rationales and 72 requirements, including 22 impacted requirements. As this dataset is publicly available, we restrict its use to the selection and training phases to prevent data leakage from affecting our results, particularly when employing LLMs in \textit{ProReFiCIA}. Therefore, we use this dataset only for RQ1 and RQ2 and to train the NLI model described in Section~\ref{sec:filtering}.
\sectopic{\projNameNew.}
This dataset is derived from a satellite data link management system. It consists of 5 change rationales and 192 requirements, including 27 impacted requirements. This dataset is also used in RQ1 and RQ2 and to train the NLI model, as it contains the same number of change rationales as WASP, making the selection and training phases more robust to changes in dataset characteristics.
\sectopic{\projName.}
This dataset is derived from the same domain and system as \projNameNew~. It includes 11 change rationales and 33 impacted requirements. This subset is used solely as a test set in RQ3, RQ4, and RQ5 to evaluate \textit{ProReFiCIA} in a realistic manner, with and without domain knowledge, and to compare it against baselines.

All datasets contain fewer than half of their requirements that are identified as impacted by the corresponding change rationales. This suggests that, in practice, only a relatively small proportion of the overall requirements is typically affected by a given change. Consequently, change impact analysis must be highly accurate to capture these critical but limited subsets.
The \projNameNew~ and \projName~ datasets are more complex than the WASP dataset, as they contain more requirements, with fewer than 17\% affected by any of the given change rationales. Moreover, the change rationales are more implicit and indirect, use fewer shared terms with the requirements they impact, and rely more on contextual cues. In contrast, the WASP dataset exhibits a higher proportion of impacted requirements (31\%). Additionally, the change rationales in this dataset are generally more explicit and direct than those in \projName~, often reusing terminology that closely aligns with the impacted requirements. This disparity indicates that change impact analysis for \projNameNew~ and \projName~ are inherently more challenging: a much larger pool of requirements must be searched to identify a comparatively smaller fraction that is actually relevant. As a result, the risk of overlooking impacted requirements is higher, and retrieval precision becomes particularly critical. Therefore, using WASP and \projNameNew~ as training datasets, spanning different domains and levels of complexity, enables a more comprehensive evaluation of prompt variations across LLMs and helps select the best model-prompt, as explained in RQ1 and RQ2. Furthermore, using the unseen, complex dataset \projName~ as the test set for RQ3, RQ4, and RQ5 provides a more realistic assessment of \textit{ProReFiCIA} under challenging conditions, without the risk of data leakage.

\begin{table*}
\caption{Characteristics of Selected LLM Models}
    \centering
\label{tab:llms_info}
\begin{tabularx}{\textwidth}{@{} p{0.09\textwidth} @{\hskip 1em} p{0.25\textwidth} @{\hskip 0.5em} p{0.1\textwidth} @{\hskip 1em} *{4}{>{\centering\arraybackslash}X}@{}}
    \toprule
    \textbf{Model} & \textbf{Version} &  \textbf{\#Parameters} & \textbf{Max Input} & \textbf{Size} & \textbf{O/C Source} \\
    \midrule
    \textbf{GPT-4o} & GPT4-omni  & unknown & 128K Tokens & unknown & close \\ \\
    \textbf{Deepseek} & deepseek-R1 & 671B & 1M Tokens & 400GB &  open \\ \\
    \textbf{LLaMa} & llama3-405B & 405B & 128K Tokens & 243GB &  open \\ \\
    \textbf{Mistral} & mistral-large-latest & 123B & 128K Tokens & 73GB & open \\ \\
    \textbf{Gemini} & gemini-2.0-flash-lite & unknown & 1M Tokens & unknown & close  \\
    \bottomrule
\end{tabularx}
\end{table*} 
\subsection{LLM Model Selection}\label{model-selection}
As part of our proposed approach, which relies on prompt-based techniques, we need to select an appropriate LLM to execute the designed prompts. 
Different LLMs offer distinct capabilities, training foundations, and optimization strategies, all of which can influence how effectively they support requirements engineering tasks. 
Relying solely on a single model should not be considered best practice, as the rapid obsolescence of LLMs can undermine the validity, generalizability, and reproducibility of results~\cite{zadenoori2025large}. To address this limitation, evaluating multiple LLMs can mitigate model-specific bias and provide a more robust understanding of how findings generalize across different LLM architectures.

To provide a comprehensive evaluation, we therefore selected five widely used and recently impactful models in this domain: GPT~\cite{Radford:18}, Deepseek~\cite{guo2025deepseek}, LLaMa~\cite{dubey2024llama}, Mistral~\cite{jiang2023mistral7b}, and Gemini~\cite{team2023gemini,team2024gemini}.
Each model brings complementary strengths. GPT, one of the earliest and most established LLM families, has been widely adopted for RE-related tasks, making it a natural baseline for comparison. Deepseek represents a newer generation of models optimized for reasoning and efficiency, and it has shown promising results in structured tasks relevant to RE. LLaMa, developed as a scalable open-source alternative, is particularly valuable for reproducibility and community-driven experimentation, enabling fine-tuning in RE-specific contexts. Mistral, known for its lightweight yet high-performing architecture, provides an opportunity to test how smaller, efficiency-oriented models handle RE workloads. Finally, Gemini integrates multimodal reasoning capabilities and cutting-edge advances in alignment, offering insights into how next-generation LLMs can push the boundaries of requirements analysis and automation. For each model, we selected versions with more parameters, as they often retain more world knowledge and domain-specific patterns from pretraining. We also focus on models that have been extensively tested in the RE domain. The chosen versions of all LLMs are listed in Table~\ref{tab:llms_info}.

By including these five models, we aim to investigate how different LLMs perform and generalize when applied to CIA in RE. This comparative perspective lets us explore not only their individual effectiveness but also how diverse architectures and training paradigms shape outcomes in RE contexts.

\subsection{Research Questions (RQs)}~\label{subsec:RQs}
\sectopic{RQ1. Which large language models demonstrate more robust performance across different prompts and datasets?} 

The goal is to examine LLM stability across a wide range of settings. Specifically, we evaluate their performance across 64 distinct prompts and two datasets that differ in domain and complexity. 

\sectopic{RQ2. Which prompt details most influence LLMs' performance, and what is the optimal prompt?}

Our objective is to examine the sensitivity of the robust LLMs identified in RQ1 to optional prompt details. By doing so, we aim to understand how prompt design and dataset characteristics, such as the domain or complexity of the requirements, jointly influence model behavior. Moreover, we identify which details influence the model and examine how their inclusion or exclusion affects performance. Using this information, we identify the prompt configuration that achieves both reasonably strong performance and generalization across datasets.

\sectopic{RQ3. How does our proposed post-processing step in \textit{ProReFiCIA} contribute to performance improvement?} 

We apply refinement and filtering stages to improve recall and precision. These steps mitigate LLM limitations that prompt engineering alone cannot address.
\sectopic{RQ4. How does \textit{ProReFiCIA} fare against a simpler baseline solution?}  

We evaluate \textit{ProReFiCIA}'s performance against a simpler, more cost-effective baseline. For this analysis, we used a CIA baseline from the literature and applied it to the WASP dataset. Additionally, we implemented simpler techniques, including semantic similarity with pre-trained language models.

\sectopic{RQ5. How does adding domain knowledge improve \textit{ProReFiCIA}?}

Since identifying the impact of a requirement change often demands rich contextual and domain-specific knowledge, we aim to enhance the system by integrating external descriptive documents that capture the software system’s domain knowledge. We further investigate how this added domain knowledge improves \textit{ProReFiCIA}'s performance.

\subsection{Model's Robustness Evaluation (RQ1)}\label{sec:RQ1}

\sectopic{Methodology.} 
Since the goal of this RQ is to find the most robust models across data characteristics, we use two datasets, WASP and \projNameNew~, which are from different domains, contributing the same number of change rationales.
We thus select models that generalize better across different prompt variations and datasets while maintaining high accuracy.
In particular, we identify models based on their consistent performance across different prompt variations while having high performance.  Such stability increases the likelihood that the model will maintain strong performance for the CIA task when applied to unseen test data.
A total of 64 prompts, as described in Section~\ref{sec:prompt_engineering}, are constructed and applied to the five models across WASP and \projNameNew~, yielding 320 distinct results using Recall (R), Precision (P), and F2 scores.
Recall measures the proportion of impacted requirements correctly retrieved by a specific prompt and model, indicating the model’s ability to capture all relevant requirements. Precision quantifies the proportion of retrieved requirements that are actually relevant, reflecting the model’s predictive reliability. The F2 score combines recall and precision, placing greater emphasis on recall, and provides a single metric that balances the need to retrieve as many impacted requirements as possible while maintaining a reasonable level of precision. In the context of the CIA task, the effort required to handle false positives is relatively minor compared to the risk of missing an impacted requirement, which justifies using F2.

In RQ1, the primary goal is to investigate the significance of changes in the models' outputs across 64 prompts and two datasets. 
We apply all prompt variations over all the change rationales in each dataset. Then, we use the results to assess the robustness of the models, which we define as the stability and consistency of model performance across variations in prompts and datasets.
The distribution of scores can be used to assess both the central tendency and variability of model performance. Based on this information, we can identify which models maintain more stable outputs under varying conditions and which are more sensitive to prompt or dataset changes. Therefore, we compute F2 descriptive statistics (Minimum, Maximum, Median, and Variance) for each model across all 64 prompts. 
To select models that are both high-performing and robust, we mainly focus on the variance and median of each model across both datasets, as the median reflects typical performance while remaining robust to extreme values (outliers) that may arise from prompt variations.

\begin{table*}
\centering
\caption{Statistic of 64 prompts across 5 LLMs on WASP and \projNameNew~.}
    \centering
\label{tab:box-plot-detailed}
\begin{tabularx}{\textwidth}{@{\hskip 0.5em} p{0.12\textwidth} @{\hskip 0.3em}  *{5}{>{\centering\arraybackslash}X}@{}}
    \toprule
    \textbf{Dataset} & \textbf{Model} & \textbf{Minimum} & \textbf{Maximum} & \textbf{Median} & \textbf{Variance} \\
    \midrule
    \multirow{5}{*}{WASP} & GPT-4o & 0.61 & 0.89 & 0.78 & 3e-3  \\
                               & LLaMa & 0.66 & 0.83 & 0.75 & 1e-3  \\
                               & Mistral & 0.47 & 0.77 & 0.66 & 4e-3  \\
                               & Gemini & 0.61 & 0.85 & 0.75 & 5e-3  \\
                               & Deepseek & 0.38 & 0.89 & 0.63 & 1.1e-2  \\
    \midrule
    \multirow{5}{*}{\projNameNew} & GPT-4o & 0.34 & 0.64 & 0.52 & 6e-3 \\
                               & LLaMa & 0.36 & 0.61 & 0.50 & 3e-3  \\
                               & Mistral & 0.29 & 0.67 & 0.42 & 1.2e-2  \\
                               & Gemini & 0.26 & 0.68 & 0.47 & 1.1e-2  \\
                               & Deepseek & 0.32 & 0.66 & 0.46 & 4e-3  \\
    \bottomrule
\end{tabularx}
 \end{table*}

\sectopic{Results.}
Table~\ref{tab:box-plot-detailed} reports descriptive statistics for each model's distribution across WASP and \projNameNew. The table columns-minimum, maximum, median, and variance-are calculated over the 64 prompt variations for each LLM.

On the WASP dataset, GPT-4o, LLaMa, and Gemini have the highest median value. GPT-4o and Gemini have the same maximum; however, GPT-4o has a higher median, indicating its performance is more consistent and stronger across different prompt variations than Gemini. The Mistral model has the lowest maximum performance, indicating that even under its most favorable conditions, it does not reach the peak levels achieved by the other models; however, its median is higher than Deepseek's, despite the latter's higher maximum. 
This indicates that Mistral performs more consistently compared to Deepseek. 
On the other hand, the LLaMa model not only has the highest minimum but also has the second-highest maximum and median. Among all models, LLaMa exhibits the lowest variance while maintaining a median that ranks among the top two highest. Similarly, GPT-4o obtains the top median value and the second-lowest variance, indicating both strong central performance and relatively stable behavior across settings. In contrast, Gemini and DeepSeek show the greatest variance in F2 scores across the 64 prompts, suggesting that their performance is more sensitive to prompt details.

On the \projNameNew~ dataset, the models with the highest median performance are GPT-4o and LLaMa. The LLaMa model achieves the highest minimum value, the second-highest median, and the lowest variance, indicating that it is not only the most robust but is also accurate. Mistral has the lowest minimum and median values, indicating that its performance is generally weaker than the other models across the evaluated settings. DeepSeek and Gemini achieve the highest maximum values among all models; however, their lower medians suggest that these high F2 scores are rare rather than common across prompt variations. Similar to WASP, LLaMa has the lowest variance. We observe that on \projNameNew~, DeepSeek has the second-lowest variance, whereas on WASP it exhibits the highest. A similar pattern is seen for Gemini and Mistral, whose variances also shift noticeably across the two datasets. This suggests that the stability of these models depends heavily on the characteristics of the dataset being evaluated. LLaMA and GPT-4o, however, maintain relatively stable variance across datasets, which indicates that their performance is less influenced by changes in the underlying input data.

GPT-4o and LLaMa are identified as the most robust models across both datasets, as they not only exhibit stable variance but also achieve a high median, indicating strong accuracy.

\begin{tcolorbox}[arc=1mm,width=\columnwidth,
                  top=0mm,left=0mm,  right=0mm, bottom=0mm,
                  boxrule=1pt, colback=violet!15!white,colframe=white]
\textbf{The answer to RQ1} is that, across prompts and datasets, the GPT-4o and LLaMa model exhibits a high and stable accuracy compared to the other LLMs. In contrast, on average, the Gemini and Deepseek models occasionally attain higher F2 scores but exhibit lower stability. Moreover, the Deepseek and Mistral models achieve a lower median score across datasets, suggesting weaker overall accuracy. 
\end{tcolorbox}
 
\subsection{Prompt Details Influence on LLMs Performance and Best Combination (RQ2)}\label{sec:RQ2}

\sectopic{Methodology.} 
The main goal of this RQ is to find a model-prompt variation that performs well on both WASP and \projNameNew~. To do this, we first analyze how different prompt details affect the models' F2 score when they are included using a Gradient Boosting classifier (Part A). Based on this analysis, we use a procedure to determine which details should be included in the prompt (Part B).

\textbf{Part A:} To evaluate the sensitivity of models to prompt details, we consider prompt details combinations as a set of six features: details \circled{1} to ~\circled{7}, excluding the mandatory detail \colorcircled{2}. Each combination is applied to every change rationale in the dataset, and the resulting output changes are assessed using the F2 score, which gauges sensitivity to prompt variation. These results are then analyzed using a gradient-boosting decision tree model~\cite{friedman2001greedy} to explain how prompt details determine F2. 
Decision trees perform feature selection by choosing, at each split, the feature that most reduces prediction error~\cite{friedman2001greedy}. In this context, features are details in the prompt. As the tree grows, features (details) that consistently improve predictions (F2 score) are selected more often and appear higher in the structure. In ensemble methods like gradient boosting, this process is repeated across many trees, producing more robust estimates of feature importance. These importance scores can then be used to identify and retain the most influential features. 
In gradient boosting, each new tree is trained to reduce the errors of the previous ones by following the gradient of the loss function, which in this case is the Mean Squared Error (MSE). This iterative process allows the model to capture complex relationships and feature interactions that a single decision tree would likely miss.
For feature importance, gradient boosting leverages the fact that each decision tree splits the data based on features that reduce prediction error. By aggregating these reductions across all trees, the algorithm estimates the contribution of each detail to model accuracy. In other words, importance scores are averaged across all decision trees created by gradient boosting and then normalized. The importance score calculation, for detail \textit{d} in a single decision tree, is given below:

\begin{equation}
Imp\ (d) = \frac{1}{N_{root}} \sum_{j\ \text{splits on}\ d} \Big( N_j I_j - N_{j_L} I_{j_L} - N_{j_R} I_{j_R} \Big)    
\end{equation}

\text{where:}
\begin{align*}
d &\text{ = detail \textit{d}} \\
N_j &\text{ = number of samples at node } j \\
I_j &\text{ = MSE at node } j \\
j_L, j_R &\text{ = left and right child nodes of } j
\end{align*}

Moreover, to assess whether including or excluding a specific detail improves overall model performance, we use SHAP (SHapley Additive exPlanations)~\cite{lundberg2017unified}. SHAP explains model predictions by assigning each feature a contribution value indicating its effect on the output. For gradient boosting models, it applies an efficient algorithm called TreeSHAP, which exploits the structure of decision trees to compute exact feature contributions.

  \begin{table*}
\centering
\caption{Details Importance using Gradient Boosting on LLaMa and GPT-4o.
}
\label{tab:detail-importance2}
\begin{tabularx}{\textwidth}{@{\hskip 0.5em} p{0.10\textwidth} @{\hskip 0.5em} p{0.12\textwidth} @{\hskip 0.5em} p{0.15\textwidth} @{\hskip 1em} *{6}{>{\centering\arraybackslash}X}@{}}
    \toprule
    \multirow{2}{*}{\textbf{Model}} & \multirow{2}{*}{\textbf{Dataset}} & & \multicolumn{6}{c}{\textbf{details}} \\
    \cmidrule{4-9} 
    & & & \textbf{\circled{1}} & \textbf{\circled{3}} & \textbf{\circled{4}} & \textbf{\circled{5}} & \textbf{\circled{6}} & \textbf{\circled{7}} \\
    \midrule
    \multirow{6}{*}{\textbf{LLaMa}} &
    \multirow{3}{*}{\textbf{WASP}}  & Importance Score & 0.03 & 0.36 & 0.10 & 0.20 & 0.25 & 0.06 \\
                                    && Effect & + & - & - & + & + & -\\
                                    && Rank & \nth{6} & \nth{1} & \nth{4} & \nth{3} & \nth{2} & \nth{5} \\
    \cmidrule{2-9} 
    &\multirow{3}{*}{\textbf{\projNameNew}}  & Importance Score & 0.22 & 0.09 & 0.26 & 0.03 & 0.18 & 0.21 \\
                                    && Effect & + & + & + & + & + & - \\
                                    && Rank & \nth{2} & \nth{5} & \nth{1} & \nth{6} & \nth{4} & \nth{3}  \\
    \midrule
    \multirow{6}{*}{\textbf{GPT-4o}} &
    \multirow{3}{*}{\textbf{WASP}}  & Importance Score & 0.02 & 0.56 & 0.03 & 0.12 & 0.23 & 0.04 \\
                                    && Effect & + & + & - & + & + & -\\
                                    && Rank & \nth{6} & \nth{1} & \nth{5} & \nth{3} & \nth{2} & \nth{4} \\
    \cmidrule{2-9} 
    &\multirow{3}{*}{\textbf{\projNameNew}}  & Importance Score & 0.04 & 0.38 & 0.05 & 0.32 & 0.19 & 0.02 \\
                                    && Effect & - & + & + & + & + & - \\
                                    && Rank & \nth{5} & \nth{1} & \nth{4} & \nth{2} & \nth{3} & \nth{6}  \\
\bottomrule
\end{tabularx}
\end{table*}

\textbf{Part B:} To identify a prompt combination that performs well across both datasets, we employ the following procedure: a detail is included in the prompt if it enhances the model based on its effect on the model’s output.
Since the effect of details on a model may not be entirely consistent across the two datasets, we focus on analyzing whether they provide an overall gain. In other words, when a detail’s effect is consistent across both datasets, we consider it a stable selection signal: a positive effect leads to inclusion in the prompt, and a negative effect leads to exclusion. If a detail’s effect is inconsistent across datasets, we consider the effect value for the dataset with the higher importance for that detail. This selection ensures that if the inclusion or exclusion of a detail improves performance on only one dataset, its negative effect on the other dataset is relatively small.
Moreover, since we have two models and are looking for a prompt combination that produces the same effect across the two datasets, we select the model with the lower inconsistency in effect values across the two datasets. Therefore, this procedure enables the selection of a model that demonstrates the highest robustness across datasets, together with a prompt variation that captures generally effective patterns rather than being tuned to a single dataset, allowing for a more reliable evaluation of how the configuration may generalize to unseen datasets, which we investigate in Section~\ref{sec:RQ3}.

\sectopic{Results.}
Table~\ref{tab:detail-importance2} reports on the importance of details in the prompt presented in Figure~\ref{fig:prompts-flow} (\circled{1} for role component, \colorcircled{2}, \circled{3}, and \circled{4} for instruction component, and \circled{5}, \circled{6}, and \circled{7} for context component), providing their importance scores and effect across WASP and \projName~ for both LLaMa and GPT-4o.
Our primary focus is on the top three details (those with the highest importance scores), as their higher scores indicate the strongest influence on the model’s behavior. The remaining details receive relatively low scores, indicating that their impact on the model’s predictions is small and unlikely to materially alter its output.
The results in Table~\ref{tab:detail-importance2} indicate that the importance ranking is not entirely consistent across both models. However, GPT-4o exhibits a more consistent pattern across both datasets compared to LLaMa. For instance, for GPT-4o, details \circled{3}, \circled{5}, and \circled{6} consistently rank among the top three most influential factors affecting performance. In contrast, for LLaMa, the rankings of the top three most influential details are inconsistent across datasets.
Detail \circled{1} (the role) has a small effect on GPT-4o performance across both datasets. However, it becomes more effective when applied to LLaMa and on the more complex dataset \projNameNew~. Details \circled{3} and \circled{4}, however, are mainly effective on \projNameNew~, across both models. We observe that these details encourage the model to retrieve more requirements, as the change rationales in this dataset are more implicit, and the model tends to overlook the relationships to the provided requirements. These details in \projNameNew~ helped retrieve relevant requirements even when the change rationale is expressed indirectly or when requirements are not explicitly stated in the same terminology as the change rationale. However, these details also led to a higher number of false positives, indicating reduced precision in simpler or more explicit datasets like WASP. Details \circled{5} and \circled{6} appear to consistently have a positive effect across both models and datasets. We observe that this detail helps guide the models' retrieval process toward more accurate selections by improving the model’s understanding of the task context. Finally, detail \circled{7} shows a consistently negative effect on both models. We observe that this detail slightly reduces models' performance, as indicated by its low importance score. This may be because it provides domain information at an abstract level, which can confuse the model.

To identify the optimal prompt combination using the procedure described in the methodology, we exclude the LLaMA model at this stage, as the effects of details are less consistent than for GPT-4o. Following the procedure, details \circled{3}, \circled{5}, \circled{6}, and \circled{7} have the same effect. Therefore, using the GPT-4o model details \circled{3}, \circled{5}, and \circled{6} are included in the prompt, and detail \circled{7} is excluded. For details \circled{1} and \circled{4}, as there is an inconsistency, we consider the effect value of the one that has the higher importance score. Therefore, we decide not to include detail \circled{1} as the importance score of the negative effect on WASP (0.04) is more than the positive effect on \projNameNew~ dataset (0.01). Moreover, for detail \circled{4} we include it as its positives effect on \projNameNew~ is higher than its negative effect on WASP.
This will give use the prompt \colorcircled{2}\circled{3}\circled{4}\circled{5}\circled{6} which is \texttt{P53}. Therefore, GPT-4o using \texttt{P53} is selected as the best model–prompt combination. Our analysis shows that this configuration ranks in the top 10\% on each dataset.

\begin{tcolorbox}[arc=1mm,width=\columnwidth,
                  top=0mm,left=0mm,  right=0mm, bottom=0mm,
                  boxrule=1pt, colback=violet!15!white,colframe=white]
\textbf{Answer to RQ2:} The \texttt{P53} prompt with the GPT-4o model is the most promising option across both datasets. GPT-4o shows greater consistency in the details effect across two datasets, with the highest sensitivity (higher importance score) to context-related details. However, LLaMa shows greater sensitivity to details across datasets, with the impact of instruction-related details varying by dataset. In particular, these details tend to have a positive effect on the more implicit and complex dataset, whereas they have a negative impact on the more explicit dataset.
\end{tcolorbox}

\subsection{Refinement and Filtering in \textit{ProReFiCIA} (RQ3)}\label{sec:RQ3}
\sectopic{Methodology.}
The primary objective of this RQ is to examine the effectiveness of our proposed post-processing steps in \textit{ProReFiCIA}. We use the optimal model–prompt combination identified in RQ2, \texttt{P53} with GPT-4o, to assess how much the proposed post-processing improves prompt results. 
In this RQ, we evaluate this combination on the unseen dataset \projName~.
To this end, we compare performance before and after applying the post-processing steps. 
We evaluate performance based on how well the predicted impact set includes all impacted requirements while minimizing the analyst’s effort to filter out false positives. Therefore, we use two metrics: \textit{Effectiveness}, which assesses how well the retrieved impact set contains all the impacted requirements, and \textit{Cost}, which evaluates the effort required for validation. For effectiveness, we use recall (Equation~\ref{eff}); for cost, we compute the average proportion of requirements retrieved relative to the total number of requirements in the dataset (Equation~\ref{cost}). 
\begin{equation}\label{eff}
    eff = \frac{1}{N_c} \sum_{i=1}^{N_c} \frac{TP_{c_i}}{TP_{c_i} + FN_{c_i}}
\end{equation}
\begin{equation}\label{cost}
    cost = \frac{1}{N_c} \sum_{i=1}^{N_c} \frac{TP_{c_i} + FP_{c_i}}{N_{req}}
\end{equation}
$N_c$ represents the number of change rationales in the dataset, $N_{req}$ the total number of requirements in the dataset, $TP_{c_i}$ and $FP_{c_i}$ represent the number of true positives and false positives retrieved in the impact set for the change rationale $i$. 
A higher effectiveness value indicates a more complete impact set, with a maximum possible value of 1 (100\%). Conversely, a lower cost corresponds to a more efficient impact set.

The post-processing refinement step enhances effectiveness by increasing the number of impacted requirements correctly retrieved. In contrast, the filtering step reduces analysis costs by minimizing the number of requirements that need to be reviewed in the final impact set.
In the filtering step, we fine-tune the NLI model using the training set processed with GPT-4o-\texttt{P53}, which provides the model's rationales used as NLI inputs, as described in Section~\ref{sec:filtering}. Because of the limited number of labeled instances in our context, we freeze all model layers during fine-tuning except the classifier layer (the last layer). The small number of tuned parameters lets us leverage the contextual knowledge encoded in the NLI model while adapting the classifier to our CIA task, which typically involves limited data.

\begin{table*}
\centering
\caption{\textit{ProReFiCIA} stages performance
}
\label{tab:reficia}
\centering
\begin{tabularx}{\textwidth}{@{\hskip 0.5em} p{0.15\textwidth} @{\hskip 0.5em} p{0.2\textwidth} @{\hskip 0.5em} *{5}{>{\centering\arraybackslash}X}@{}}
    \toprule
    & &\multicolumn{5}{c}{\textit{\projName}} \\
    \cmidrule(lr){3-7}
    \textbf{Setting} & \textbf{Stage} & \textbf{TP} & \textbf{FN} & \textbf{FP} & \textbf{eff} & \textbf{cost} \\
    \midrule
    \multirow{3}{*}{\textbf{GPT-4o-\texttt{P53}}}  & w/o  & 23 & 10 & 9 & 73.3\% & 1.5\%  \\
        & Refinement  & 28 & 5 & 49 & 85.7\% & 3.6\%  \\
        & Refinement+Filtering  & 28 & 5 & 30 & 85.7\% & 3.0\%  \\
    \bottomrule
\end{tabularx}
    \begin{tablenotes}
     \item \textit{w/o: without any post-processing, P: precision, eff: effectiveness} 
    \end{tablenotes}
\end{table*} 
\sectopic{Results.}
Table~\ref{tab:reficia} presents the results on \projName~. The performance changes are reported across the post-processing stages, starting with refinement and then filtering.

As reported in the table, the refinement step improves effectiveness. However, it also increases cost, which the filtering step addresses by reducing false positives in the impact set for each change rationale.
GPT-4o-\texttt{P53} without post-processing achieved an effectiveness of 73.3\% on the \projName~dataset.
Furthermore, the model exhibits a low cost of under 2\%, indicating strong overall cost-effectiveness. Nonetheless, to further enhance the results, we apply additional refinement and filtering steps. The refinement step improved effectiveness by about 12 pp and identified five additional impacted requirements that the initial prompt execution missed. Therefore, the refinement step is effective, as it further mitigates the lost-in-the-middle problem described in Section~\ref{sec:refinement}, where the LLM tends to pay less attention to requirements midway through the prompt. However, this stage also increased costs to 3.6\%. 
The filtering step, by contrast, reduced cost by 0.6 pp compared to the refinement step. The filtering step uses both the NLI model and ranking positions, each helping preserve true positives while reducing false negatives. Based on our analysis, the NLI model and the ranking component, when applied separately, can eliminate 65\% and 57\% of the false positives, respectively. Moreover, when applied separately, the NLI model and the ranking component preserve 65\% and 91\% of the true positives, respectively. However, combining these two techniques reduces false positives by 39\% while preserving 100\% of the true positives.
Based on these results, the analyst can, on average, identify 85.7\% of the impacted requirements by examining just 3\% of the total set of requirements.

To further investigate the sensitivity of the filtering step, we experiment with different proportions of the ranked list used during filtering, as described in Section~\ref{sec:filtering}. Table~\ref{tab:filtering-sens} shows the performance of \textit{ProReFiCIA} when 25\% and 75\% of the ranked list are used in the filtering step. Using a smaller proportion reduces FPs but also removes TPs.
On the other hand, using 75\% of the ranked list results in more FPs but also ensures more true positives are covered.

\begin{table*}

\caption{Sensitivity of the filtering step to different proportions of the ranked list
}
\label{tab:filtering-sens}
\centering
\begin{tabularx}{\textwidth}{@{\hskip 0.5em} p{0.15\textwidth} @{\hskip 0.5em} p{0.15\textwidth} @{\hskip 0.5em} *{5}{>{\centering\arraybackslash}X}@{}}
    \toprule
    & & \multicolumn{5}{c}{\textit{\projName}} \\
    \cmidrule(lr){3-7}
    \textbf{Combination} & \textbf{Setting} & \textbf{TP} & \textbf{FN} & \textbf{FP} & \textbf{eff} & \textbf{cost} \\
    \midrule
    \multirow{3}{*}{\textbf{GPT-4o-\texttt{P53}}}  & Filtering (50\%) & 28 & 5 & 30 & 85.7\% & 3.0\%  \\
    \cmidrule{2-7}
                                                  & Filtering (25\%) & 24 & 9 & 26 & 77.7\% & 2.6\% \\
                                                  & Filtering (75\%) & 28 & 5 & 36 & 85.7\% & 3.3\% \\
    \bottomrule
\end{tabularx}

 \end{table*} 
\begin{table*}
\caption{Comparison between the iterative approach and \textit{ProReFiCIA}, where the full set of requirements is provided in the prompt, as well as a setting using a batch size of 100.
}
\label{tab:iterative}
\centering
\begin{tabularx}{\textwidth}{@{\hskip 0.5em} p{0.15\textwidth} @{\hskip 0.2em} p{0.18\textwidth} @{\hskip 0.5em} *{5}{>{\centering\arraybackslash}X}@{}}
    \toprule
    & & \multicolumn{5}{c}{\textit{\projName}} \\
    \cmidrule(lr){3-7}
    \textbf{Combination} & \textbf{Setting} & \textbf{TP} & \textbf{FN} & \textbf{FP} & \textbf{eff} & \textbf{cost} \\
    \midrule
    \multirow{1}{*}{\textbf{GPT-4o-\texttt{P53}}} & Iterative & 33 & 0 & 252 & 100\% & 35.9\%  \\
    \multirow{1}{*}{\textbf{GPT-4o-\texttt{P53}}} & \textit{$\text{ProReFiCIA}_{batch=100}$} & 30 & 3 & 76 & 95.8\% & 5.3\%  \\
    \multirow{1}{*}{\textbf{GPT-4o-\texttt{P53}}} & \textit{ProReFiCIA} & 28 & 5 & 30 & 85.7\% & 3.0\%  \\
\bottomrule
\end{tabularx}
\end{table*} 

To more effectively assess the performance of the CAG technique used in the prompt, with respect to varying numbers of requirements, we compare iterative querying against \textit{ProReFiCIA} in a setting where GPT-4o-\texttt{P53} is applied to the full set or to batches of 100 requirements, where for each batch the whole process in \textit{ProReFiCIA} is applied independently and the final outputs are aggregated. This batching strategy simulates a scenario in which the entire set of requirements cannot be inserted within a single prompt execution.
Table~\ref{tab:iterative} shows the results using these settings. When applied iteratively without post-processing (which is not feasible here, since the model receives only one requirement at a time), the cost exceeds 35\%, while precision drops to 11.5\%. This suggests that for each change rationale, the model retrieves many irrelevant candidate links, significantly increasing the validation burden.
However, when \textit{ProReFiCIA} is applied with a batch size of 100 requirements, the number of false positives decreases to 76. This indicates that using CAG and post-processing effectively reduces the number of retrieved links. This behavior also occurs when the model receives the full set of requirements. Therefore, as more requirements are included, the model can leverage richer shared context across requirements to make more informed, selective decisions about which requirements are truly impacted. As a result, it reduces unnecessary retrievals while maintaining comparable effectiveness, indicating that the model benefits from contextual signals present in larger batches when identifying relevant impacted requirements. Based on these results, we suggest applying the refinement stage with a large number of requirements (>100) in the prompt.

\begin{figure}
    \centering
    \includegraphics[width=0.5\linewidth]{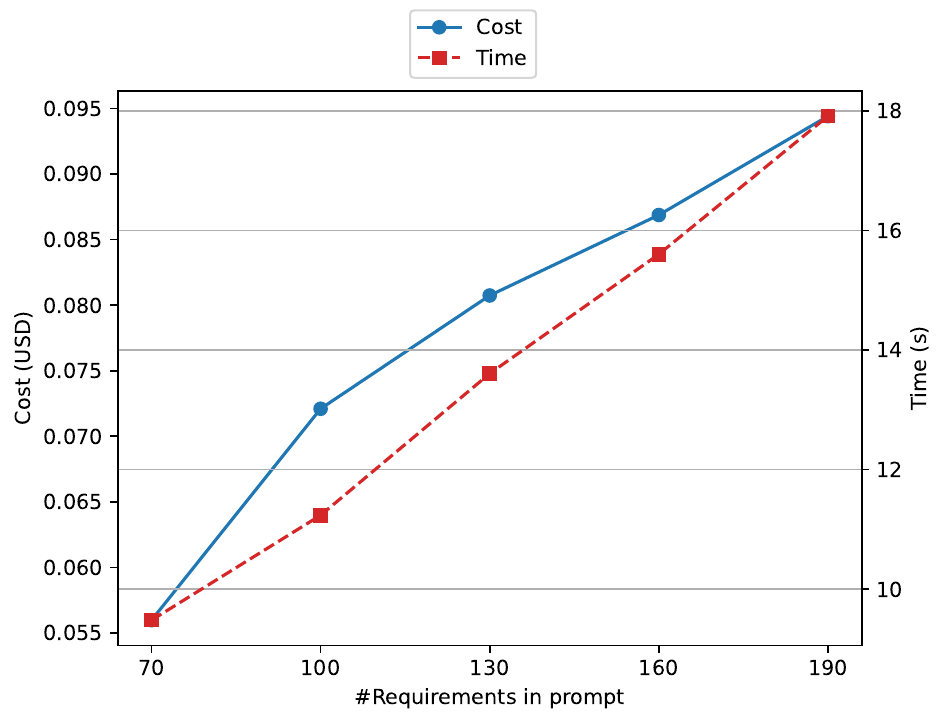}
    \caption{Scalability analysis of average inference cost and runtime across prompt-dependent stages of \textit{ProReFiCIA} as a function of the number of input requirements.}
    \label{fig:scalability}
\end{figure}

Finally, to assess both the inference cost and runtime efficiency of \textit{ProReFiCIA}, we measure prompt latency across the initial, refinement, and sorting steps, and analyze how computational time and cost scale with the number of requirements in the prompt. Figure~\ref{fig:scalability} presents the average scalability trends over multiple runs for a single batch. As shown, both cost (tokens) and runtime grow approximately linearly with the number of requirements, indicating that the method scales proportionally and remains computationally stable as input size increases.

\sectopic{Statistical Significance.}
Since the optimal model-prompt combination already performs well in the CAG setting (with all requirements) before applying the post-processing steps, the sample sizes in Table~\ref{tab:reficia} are too small to assess the statistical significance of differences in TPs, FNs, and FPs across the individual post-processing steps. The results, therefore, need to be confirmed on larger datasets. However, we present insights into the benefits of post-processing steps, and we also present qualitative findings from a manual analysis of false positives and false negatives in the error analysis below. Further, as discussed earlier, even a small number of additional false positives may matter in our application context, since our priority is not to miss any impacted requirements. 

To assess statistical significance between \textit{ProReFiCIA} and the setting that uses iterative querying (one requirement at a time) of the same model-prompt combination without post-processing steps (Table~\ref{tab:iterative}), we apply the Fisher exact test~\cite{fisher1922interpretation} to test differences in outcome proportions between the two settings, under the null hypothesis that the proportions are identical. We use this test to compare true positives versus false positives, as well as true positives versus false negatives. The p-values of the differences in proportions between \textit{ProReFiCIA} and the iterative setting are 1e-4 and 5.3e-2, respectively, for the two proportions. Thus, the proportions of true positives versus false positives yield a p-value below commonly used significance thresholds ($\alpha < 0.05$). For the proportion of true positives versus false negatives, the samples are again very small; as a result, the p-value is borderline. 
Moreover, the same test using \textit{ProReFiCIA} with different batch sizes (all requirements and a batch size of 100) yields p-values of 1.6e-2 and 7e-1, respectively, indicating a significant improvement in true positives over false positives when all requirements are provided to the model in a single prompt. Again, due to the small sample size, the difference in proportions between true positives and false negatives is not significant.

\sectopic{Error Analysis.}
In this section, we analyze the sources of errors. 
False negatives occur when the model fails to identify a valid link between a requirement and corresponding change rationale. We observe that the refinement step successfully recovered only 5 of the 10 false negatives. These cases occur because they require more domain knowledge. For example, one change rationale cites the Electromagnetic Compatibility (EMC) guideline, which specifies requirements to ensure the proper operation of electronic systems in their electromagnetic environment. For this change rationale, the model failed to retrieve requirements related to environmental conditions with no direct mention of the EMC guideline. This suggests that the model struggles to capture the domain-specific connections between the two artifacts. Therefore, successfully retrieving these missed requirements often depends on a deeper domain understanding, including the specific terminology and contextual details used in both the requirement descriptions and the change rationales. Without sufficient domain knowledge, the LLM may overlook relevant semantic cues or misinterpret specialized language. Therefore, capturing such links requires providing the model with more detailed domain knowledge, which we investigate further in RQ5 in section~\ref{sec:RQ5}.
Moreover, our in-depth analysis of false positives reveals that the LLM frequently relies on correlational signals between the change rationale and the requirement to determine impact, rather than performing true causal reasoning. In particular, the model tends to focus on surface-level cues or co-occurrence patterns—such as shared terminology—rather than assessing whether a functional dependency exists. This behavior aligns with prior findings that LLMs struggle to distinguish causation from correlation~\cite{jin2024can}.  
To mitigate such false positives during generation, further investigation of LLMs' reasoning patterns is needed to understand how they reach these conclusions. However, this analysis is left for future work.

\begin{tcolorbox}[arc=1mm,width=\columnwidth,
                  top=0mm,left=0mm,  right=0mm, bottom=0mm,
                  boxrule=1pt, colback=violet!15!white,colframe=white] 
\textbf{Answer to RQ3:} The post-processing steps in \textit{ProReFiCIA} increase the cost by 1.5 pp, while it also improves effectiveness by up to 12 pp. This indicates that the additional overhead is relatively small compared to the potential gains in effectiveness, suggesting a favorable trade-off. 
\end{tcolorbox}

\subsection{ProReFiCIA comparison to Baselines (RQ4)}\label{sec:RQ4}

\sectopic{Methodology.}
We evaluate our proposed technique, \textit{ProReFiCIA}, against baseline methods by comparing their performance using the effectiveness and cost metrics, as discussed in RQ4. We use two baselines: (1) $CoT$, which leverages LLM combined with RAG and (2) the semantic similarity technique which employs LLM embeddings with three different cutoff point selection methods: a fixed threshold ($S_{T1}$), the cutoff point selection approach proposed in NARCIA~\cite{Arora:15Tool} ($S_{T3}$), and a dynamic selection method ($S_{T2}$). These details are explained in Section~\ref{baseline}.

\sectopic{Results.}
Table~\ref{tab:baselines} lists, for each approach, the number of TPs, FPs, and FNs along with their effectiveness and cost.

$S_{T1}$ has the highest cost, indicating that, on average, it fails to distinguish between impacted and unimpacted requirements by retrieving all requirements for a given change rationale. Recall from Section~\ref{baseline} that $T1$ uses a constant threshold of 0.5. This observation suggests that, in most cases, the LLM embeddings capture a strong semantic relationship between a requirement and a rationale, reflecting a generally high degree of similarity. However, the requirements are not always directly affected, as semantic overlap often reflects shared concepts or related contexts rather than an actual change dependency. The high semantic similarity may occur because the change rationale and the requirements share a similar context, leading to misleading similarity scores based on surface-level overlap rather than a true, reasoning-based relation between a change rationale and a requirement.
On the other hand, $S_{T2}$ appears to be more conservative than $S_{T1}$. $S_{T2}$ demonstrates the lowest cost, which indicates that $S_{T2}$ tends to have a lower retrieval rate. Recall from Section~\ref{baseline} that $T2$ uses a cutoff point where the maximum difference in scores occurs. We observe that this technique leads to more balanced behavior than $S_{T1}$, achieving effectiveness above 51.7\% while retrieving only 2.2\% of the requirements. This finding suggests that the method performs more effectively when the difference between similarity scores is used rather than their absolute values. 
$S_{T3}$ achieves higher effectiveness than $S_{T2}$ by relaxing the selection criteria to choose the last significant peak as the cutoff point, defined as the difference between the scores in a sorted list that exceeds one-third of the maximum peak, rather than relying solely on the highest peak as the cutoff point. However, this comes at a cost, as the number of retrieved requirements increases under the broader selection strategy. Similar to $S_{T2}$, this method achieves a better balance between effectiveness and cost than $S_{T1}$ by leveraging differences in similarity scores in its selection algorithm.

$CoT$ also demonstrates a better balance between effectiveness and cost. We can observe that $CoT_{gpt4o}$'s performance is comparable to $S_{T2}$, though it produces more false positives. It is worth noting that in the \projName~dataset, 9.1\% of the requirements are lost during the RAG (retrieval step), meaning their semantic similarity scores did not rank them among the top results.

\textit{ProReFiCIA}, on the other hand, achieves high effectiveness while incurring low cost. \textit{ProReFiCIA} outperforms both $S_{T2}$ with 34 pp higher effectiveness, and $S_{T1}$ with 96 pp lower cost. Moreover, \textit{ProReFiCIA} also outperforms $CoT$ on both effectiveness and cost.
To assess statistical significance, similar to RQ3, we apply Fisher's exact test to compare outcome proportions between the two techniques, under the null hypothesis that the proportions are identical. We use this test to compare the proportions of true positives and false positives, as well as true positives and false negatives, between \textit{ProReFiCIA} and the baselines. This allows us to assess two key aspects of performance: (1) how accurately the model retrieves impacted requirements, and (2) how effective it is at not missing impacted requirements. This test evaluates the statistical significance of \textit{ProReFiCIA}’s performance relative to each baseline, specifically its ability to maximize true-positive retrieval while maintaining a low number of false positives. The p-values of the differences in proportions between \textit{ProReFiCIA} and baselines are $CoT_{gpt4o}$, $S_{T1}$, $S_{T3}$, and $S_{T2}$ are (1e-4, 1.6e-3), (1e-4, 4e-1), (1e-4, 2e-2), and (1e-2, 1.6e-3), respectively for the two proportions. For all methods, at least one of the two compared proportions yields a p-value below the commonly used significance threshold ($\alpha < 0.05$), rejecting the null hypothesis and providing strong statistical evidence that their performance differences are significant. Therefore, \textit{ProReFiCIA} demonstrates superior performance compared to all baselines, effectively maximizing effectiveness while minimizing cost.

 \begin{table*}
\caption{Comparison of \textit{ProReFiCIA} with baselines}
\centering
\label{tab:baselines}
\begin{tabularx}{\textwidth}{@{} p{0.1\textwidth} @{\hskip 1em} *{6}{>{\centering\arraybackslash}X}@{}}
    \toprule
    & \multicolumn{5}{c}{\textit{\projName}}  \\
    \cmidrule(lr){2-6}
    \textbf{Technique} & \textbf{TP} & \textbf{FN} & \textbf{FP} & \textbf{eff} & \textbf{cost} \\
    \midrule
\textbf{$CoT_{gpt4o}$} &  15 & 18 & 156 & 35.1\% & 8.1\%  \\
    $S_{T1}$ & 31 & 2 & 2072 & 86.3\% & 99.0\% \\
    $S_{T2}$ & 15 & 18 & 32 & 51.7\% & 2.2\% \\
    $S_{T3}$ & 19 & 14 & 785 & 66.8\% & 38.1\% \\
    \midrule
    \textbf{\textit{ProReFiCIA}} & 28 & 5 & 30 & 85.7\% & 3.0\% \\
    \bottomrule
\end{tabularx}
 \end{table*}  
\begin{tcolorbox}[arc=1mm,width=\columnwidth,
                  top=0mm,left=0mm,  right=0mm, bottom=0mm,
                  boxrule=1pt, colback=violet!15!white,colframe=white]
\textbf{Answer to RQ4:} \textit{ProReFiCIA} achieves significantly better performance in effectiveness and cost than the baselines. These results suggest that \textit{ProReFiCIA} provides a more balanced trade-off between effectiveness and cost for the CIA task, achieving both high effectiveness and significant cost savings, making it a more practical and scalable solution in real-world settings.
\end{tcolorbox}

\subsection{Addition of Domain Knowledge (RQ5)}\label{sec:RQ5}

\sectopic{Methodology.}
The main objective of this RQ is to determine whether adding more descriptive information about the system can reduce retrieval errors caused by a lack of domain knowledge. To provide domain information relevant to \projName~, we focus on a widely available source, Wikipedia pages about satellite concepts. To identify relevant concepts, we prompt our base model, GPT-4o, for each requirement, asking it to extract potential satellite-related concepts from Wikipedia and return the corresponding links. 

\begin{tcolorbox}[arc=1mm,width=\columnwidth,top=0mm,left=0mm,  right=0mm, bottom=0mm,
                  boxrule=0.3pt,
               colback=gray!15!white,
               colframe=black, breakable, 
]
               
\texttt{
Give me the links to all concepts and entities related to the Satellite domain in the following requirement that exist on Wikipedia.
\newline
Requirement: <requirement>
\newline
Your output should be Wikipedia page links, each on a separate line.
}
\end{tcolorbox}

Applying this prompt over 192 requirements yields 489 unique Wikipedia pages. Then we crawl each link and extract the first paragraph of the Wikipedia page as the concept description, and store it in our domain knowledge file.
Based on such domain knowledge, for each dataset, we apply RAG to retrieve the top-most relevant chunk for a given change rationale and incorporate it into the prompt. We use a chunk size of 500 characters with a 50-character overlap, select the top first chunk, and include it in the context provided to the model. To better evaluate the effect of adding domain knowledge, we incorporate it into the prompt at every stage where we query the model (including before and during refinement and sorting), ensuring it is available to the model throughout these phases to maximize the likelihood of retrieving true positives.

\begin{table*}
\small
\centering
\caption{Performance of \textit{ProReFiCIA} without and with RAG (domain-knowledge).
}
\label{tab:domain-knowledge}
\centering
\begin{tabularx}{\textwidth}{@{\hskip 0.5em} p{0.1\textwidth} @{\hskip 0.5em} p{0.2\textwidth} @{\hskip 0.5em} *{5}{>{\centering\arraybackslash}X}@{}}
    \toprule
    & & \multicolumn{5}{c}{\textit{\projName~}} \\
    \cmidrule(lr){3-7}
    \textbf{} & \textbf{Stage} & \textbf{TP} & \textbf{FN} & \textbf{FP} & \textbf{eff} & \textbf{cost}  \\
    \midrule
    \multirow{3}{*}{\textbf{w/o RAG}}  & w/o & 23 & 10 & 9 & 73.3\% & 1.5\%\\
                                                  & Refinement & 28 & 5 & 49 & 85.7\% & 3.6\%  \\
                                                  & Refinement+Filtering & 28 & 5 & 30 & 85.7\% & 3.0\%  \\
    \midrule
    \multirow{3}{*}{\textbf{RAG}} & w/o & 23 & 10 & 19 & 73.3\% & 1.9\% \\
                                                  & Refinement & 31 & 2 & 57 & 97.3\% & 4.2\%  \\
                                                  & Refinement+Filtering & 30 & 3 & 35 & 95.8\% & 3.4\%\\
    \bottomrule
\end{tabularx}
    \begin{tablenotes}
     \item \textit{w/o RAG: without RAG, w/o: without any post-processing, eff: effectiveness} 
    \end{tablenotes}

 \end{table*} 

\sectopic{Results.}
Table~\ref{tab:domain-knowledge} presents the performance of \textit{ProReFiCIA} with and without the addition of domain knowledge via RAG. We observe that in the first stage, the number of true positives remains unchanged, while the number of false positives increases slightly. However, during the refinement stage, when domain knowledge is used, the number of retrieved impacted requirements increases, yielding 3 additional true positives compared to the setting without domain knowledge. Based on our analysis, we note that in one of the change rationales, the phrase "EMC guideline" appears without further explanation. "EMC guideline" refers to the Electromagnetic Compatibility (EMC) guideline, which sets out requirements and best practices to ensure electronic systems operate properly in their electromagnetic environment. Before using RAG, the model failed to retrieve requirements related to environmental aspects (e.g., weather, temperature, humidity, etc.) of the location of a system's equipment when the term "EMC guideline" was not explicitly mentioned, and it mainly relied on surface-level relationships, which did not lead to the identification of the correct impacted requirements among all requirements that have been provided to the model. However, after introducing RAG, the model was provided with a description, enabling it to retrieve the actual impacted requirements more effectively. However, we still observe that 2 true positives have not been retrieved even when the domain knowledge is used. Based on our observation, these requirements address aspects of the change that are only indirectly affected at the technological level and therefore require additional knowledge of the software system's technical description to be properly captured.
Moreover, we observe that the filtering step successfully reduces the cost by approximately 1 pp compared to the refinement step. However, it also leads to the removal of one true positive. Based on our analysis, the missed true positive was ranked low in the sorted list, and the NLI model did not identify any relationship between it and the query. Therefore, based on RAG results, a requirements analyst can, on average, identify 95.8\% of the impacted requirements for a given change rationale by analyzing only 3.4\% of the entire set of requirements, whereas before using RAG, this effectiveness was 10 pp lower along with 0.4 pp lower cost.

\begin{tcolorbox}[arc=1mm,width=\columnwidth,
                  top=0mm,left=0mm,  right=0mm, bottom=0mm,
                  boxrule=1pt, colback=violet!15!white,colframe=white]
\textbf{Answer to RQ5:} Incorporating domain knowledge via RAG improves the performance of \textit{ProReFiCIA} by introducing essential contextual information not explicitly present in the change rationales or requirements, thereby enabling the model to better understand implicit concepts, disambiguate domain-specific terms, and more accurately identify the requirements that are impacted.
\end{tcolorbox}

\subsection{Implementation}\label{implementation}
\sectopic{LLMs.} 
We implement this approach using \texttt{Python} 3.10.12. For each LLM, we have used the following API libraries: \texttt{openai} (v 1.61.0), \texttt{genai} (v 1.0.0), and \texttt{mistralai} (v 1.5.0). Moreover, the hyperparameters for all the LLM types are the same: temperature = 0, seed = 16, and frequency and presence penalties = 0. For the CoT baseline, we used the same settings as reported in their paper.
\sectopic{NLI.} 
We implemented the NLI model using the Transformers library (v 4.52.3) with the \texttt{nli-deberta-v3-large} language model. The fine-tuning was carried out for 50 epochs with a weight decay of 1e-2, a training batch size of 5. 20 percent of the training set was used as the evaluation set, with an evaluation strategy set to \texttt{steps} and a batch size of 5. The metric for selecting the best model was f1 score with the setting of \texttt{pos\_label=1}, and the learning rate was 2e-4.
\sectopic{Gradient Boosting.} 
For tuning the gradient boosting model for data explainability, we used the \texttt{scikit-learn} library (v 1.6.1) with a fixed random state of 42. We used 71 and 40 estimators for GPT-4o and 97 and 44 for LLaMA on WASP and \projNameNew~, respectively. We determined the number of estimators using an elbow heuristic on MSE from 5 to 300 estimators, selecting the point where MSE reached its minimum; additional trees provided negligible improvement. We applied this procedure to both models and datasets to keep the results comparable. Moreover, we have used SHAP library (v 0.48.0) for feature direction.
\sectopic{LLM Embeddings.}
The LLM embedding model \texttt{gemini-embeddings-001} was loaded and utilized through the \texttt{vertexai} library (v 1.74.0). Moreover, the LLM embedding \texttt{text-embedding-3-large} was obtained via the OpenAI API with \texttt{openai} (v 1.61.0). The value of k for the top-k list in $CoT$ is selected (tuned) based on the test set (\projName~dataset) to obtain its optimal performance. Specifically, we perform a grid search over the range from 0 to the total number of requirements (192), with a step size of 5. We select the value of k with the highest F2 score on the test set, yielding a relatively low value of 60.

\sectopic{RAG.}
We implement RAG using the \texttt{Pinecone} library (v7.3.0). We use a chunk size of 500 characters with a 50-character overlap. For semantic similarity, we use the \texttt{multilingual-e5-large} embedding from \texttt{Pinecone} for retrieval scenarios. We provide domain knowledge to the model as the \texttt{system\_message}, while the prompt is supplied as the \texttt{user\_message}. To retrieve Wikipedia pages, we used the \texttt{requests} (v2.32.5) library to send HTTP requests to the Wikipedia web pages and obtain their raw HTML content. After fetching the pages, we post-processed them to extract and retain only the first relevant portion of each article's content, focusing on the initial section where the most general and informative overview is typically provided.

 \section{Threats to Validity}~\label{sec:threats}
\sectopic{Internal Validity.}
One internal validity threat in this paper concerns the effect of different prompt combinations on the quality of the impact set. To ensure that the observed differences in model performance are genuinely due to prompt structure and not to other external factors, the study systematically controls the experimental setup. To ensure consistency and reduce variability, the models’ parameters, such as the temperature, were kept fixed across all runs. We tested each prompt multiple times for each model and computed the final results by averaging the outcomes to provide a more stable, reliable estimate of performance. Additionally, to mitigate concerns about internal validity related to prompt details, we did not choose these elements arbitrarily. Instead, they were carefully designed and selected based on a systematic analysis of prompt engineering principles. 
This process ensured that any differences in results across prompts and LLMs genuinely reflected the influence of prompt details rather than uncontrolled confounding factors.
\sectopic{External Validity.} 
The most critical aspect of external validity lies in the generalizability of the proposed approach across datasets. In this study, we examine \textit{ProReFiCIA} across two distinct datasets, each representing a different domain and level of complexity. Though we do not have the number of requirements usually found in large systems, the relationships between change rationales and requirements are complex, most particularly in \projName, and pose a real challenge for the CIA, as illustrated by the example presented in Section~\ref{sec:RQ5}. We expect to observe similar relationships in other regulated domains with similar requirements. 

Moreover, we systematically conducted the experiments reported in RQ1 and RQ2 to identify the model–prompt pairs that show the highest consistency and, thus, generalizability across datasets.
However, further experiments on more diverse and large-scale datasets are necessary to strengthen the external validity of our study. Therefore, in practice, we suggest performing the model-prompt selection procedure for each new context and dataset.
\sectopic{Conclusion Validity.} Another important threat to the validity of this work concerns the reproducibility of the results. To mitigate this threat, we report all implementation details with high precision. Moreover, to facilitate independent verification and future research, we archive all experiments and relevant artifacts, allowing others to replicate our findings under identical conditions. \section{Discussion}~\label{sec:discussion}
The evaluation on our datasets shows that GPT-4o and LLaMa are the most stable and top-performing models in the CIA task. We also observe that adding more details to the prompt does not always improve performance. However, based on the results, the context details (\circled{5} and \circled{6}) remain effective when using a robust model, such as GPT-4o or LLaMa. In contrast, instruction details (\circled{3} and \circled{4}) proved to be more effective only when applied to a more complex dataset. The impact of role and domain details (\circled{1} and \circled{7}) remained relatively limited compared to the other details.  
Moreover, we designed a heuristic algorithm that selects the most generalizable prompt combination based on each detail's importance score and effect value. We also introduce two post-processing steps, Refinement and Filtering, to improve recall while controlling false positives.
Moreover, we showed that \textit{ProReFiCIA} outperforms all baseline methods across the datasets. Further, supplying all requirements directly within the LLM’s context window yields better performance than iteratively querying the model, using a small batch, or using RAG to reduce the candidate set for the CIA task. Candidate selection poses challenges, including potential inaccuracies in selecting relevant documents~\cite{chan2025don}. These issues are particularly critical in the context of CIA, where change descriptions often use diverse terminology compared to the requirements. Moreover, iteratively querying the model was found to increase the number of false positives. Therefore, for the CIA task, we recommend leveraging the LLM’s full context window, along with refinement and filtering steps, to improve prediction accuracy, reduce LLM queries, and minimize retrieval errors in the RAG technique.
In this study, all requirements fit within the context window of the LLMs used. However, in practice, the number of requirements may exceed the model’s context capacity. In such cases, we propose applying \textit{ProReFiCIA} multiple times, each time including as many requirements as possible within the context window, and then aggregating the results. We simulated this scenario using a batch size of 100 requirements. The results show that \textit{ProReFiCIA} remains more effective than iterative model querying. However, with smaller requirement batches, the number of false positives increases slightly compared to larger batches. 
Nonetheless, additional datasets are needed to further validate this strategy; we leave this as future work.

Once a change rationale is requested, along with the requirements, the resulting changes may propagate to other software artifacts, such as source code and test cases. Therefore, as future work, we aim to further investigate how to adapt \textit{ProReFiCIA} for other types of artifacts. In particular, we plan to apply and evaluate the same \textit{ProReFiCIA} setting used for natural language requirements for other natural language artifacts such as test case specifications and use cases, as they may pose similar challenges. However, for source code, we plan to apply \textit{ProReFiCIA} with the following modifications: 1) The evaluation should be conducted using additional LLMs that are specifically tuned for programming languages, such as Claude~\cite{claude2024}. 2) The instruction details (\colorcircled{2}, \circled{3}, and \circled{4}) in the prompt should be adapted according to the input artifact type. 3) The base NLI model should be replaced with a model that also supports programming languages, such as \texttt{CodeBERT}.
 \section{Related Work}~\label{sec:related}

CIA was first introduced and studied in 1993 by Arnold and Bohner~\cite{arnold1996software, bohner1996impact}. They noted that impact analysis involves identifying the potential effects of a change or predicting what must be modified to implement it.
The need to forecast and manage the impact of software changes increases as software systems grow larger and more complex. Software Change Impact Analysis (CIA) gathers current data on the software system to identify which components the proposed change will affect and how they interact. Based on Arnold and Bohner ~\cite{arnold1993impact}, there are two main perspectives for CIA, including software dependency analysis and traceability analysis~\cite{jayatilleke2018managing}.
Arnold et al.  described three main steps to analyze the change impacts in a system~\cite{arnold1993impact} :
\begin{enumerate}
  \item Analyze change specification and software artifacts
  \item Trace potential impacts
  \item Implement the requested changes
\end{enumerate}
\par Changes are initiated by a form of change request, which is referred to as the process of change specification and classification, which finishes the identification of the change type~\cite{jayatilleke2013method}. When addressing a change, many requirements cannot be considered independent in the SRS, as various relationships can exist between them. As a result, an action taken on one requirement may unexpectedly affect another. Therefore, we need to identify interdependencies among requirements~\cite{jayatilleke2018method}. Thus, we need to investigate how requirements depend on one another when they share no semantic or syntactic similarity. The investigated dependencies between requirements can be used to develop a predictive model to forecast the impact of changes.
\par In the last few decades, considerable research effort has focused on change impact analysis, particularly in software maintenance and evolution~\cite{jayatilleke2018systematic, alkaf2019automated, lehnert2011review, lehnert2011taxonomy}. Change impact analysis was applied to source codes~\cite{li2013survey}, to requirements traceability~\cite{hey2025requirements,rusdianto2024innovative,goknil2016rule, zhang2014investigating, li2008requirement}. Some researchers have reported the effects of requirements changes on data, architecture, and software design~\cite{yazdanshenas2012fine, von2002change}. Few researchers have studied the challenges of change verification and validation~\cite{bjarnason2014challenges} and of co-changing artifacts to better understand software artifact evolution~\cite{antoniol2005detecting}. Some approaches have attempted to automate change impact analysis. Alkaf et al.~\cite{alkaf2019automated} performed an automated CIA approach for User Requirements Notation models. Arora et al.~\cite{Arora2015} proposed an NLP-based approach to analyze the impact of changes in Natural Language requirements. Nejati et al.~\cite{nejati2016automated} proposed an approach to automatically identify the impact of requirements changes on system design when requirements and design elements are expressed using models. Jayatilleke et al.~\cite{jayatilleke2018method} presented a technique for requirements change analysis that relied on changes that arise at higher levels. Bano et al.~\cite{bano2012causes} conducted a systematic literature review of the causes of requirements change. The authors identified various causes and their frequency during the software development processes. They categorized the extracted causes of changes in requirements into two major categories: necessary and accidental. Aryani et al.~\cite{aryani2009change} proposed a methodology for analyzing change propagation in software using the domain-level behavioral model of a system. 
\par In the requirements specification, one solution to assess the effect of the change is to check the precise correspondence between the terms in the change and their potential definitions and expressions in other specifications. Change can progress across semantically related terms that are not exact matches or relevant syntactic variants. In this situation, it is appropriate to apply a relatedness measure that considers phrases~\cite{Arora:15b,Arora:15Tool}. In addition, dependencies in requirements play an essential role in analyzing change propagation~\cite{zhang2014investigating}. 
\par To identify relationships among requirements, many dependency and interdependency models have been developed to define and distinguish these relationships based on the requirements' structural and semantic properties~\cite{zhang2014investigating}. However, no empirical assessment has evaluated these dependency forms for usefulness and applicability ~\cite{zhang2014investigating}. To assess the potential effects of requirement changes on the overall system, Hassine~\cite{hassine2005change} applied both slicing and dependency analysis at the use case map level (rather than across requirements expressed in natural language). Baumer et al~\cite{baumer2010comparing} showed that semantic role labelling can be used to improve computational metaphor identification, and it might also be more effective at identifying relationships with semantic import than typed dependency parsing. 
\par Arora et al.~\cite{Arora:15b} showed that, in their approach, there is no need to define requirement dependencies in the early stage because the propagation condition can determine whether there is a correlation between a changed requirement and the others. Since all potential conditions cannot be enumerated, constructing an explicit dependency graph is difficult. Rather than utilizing typed dependencies, they used correlation rates to evaluate the impact of changes. Typed dependencies focus on syntactic structure and grammatical relations, while semantic roles emphasize conceptual and semantic structure~\cite{baumer2010comparing, de2008stanford}.
\par Arora et al.~\cite{Arora:15Tool} introduced NARCIA, which supports requirements change impact analysis by applying NLP techniques beyond basic keyword matching, focusing instead on the phrasal structure of requirements. Their key idea is that when a requirement changes, not all related requirements are equally affected; instead, change propagation depends on specific conditions. An analyst should specify change propagation conditions using Boolean logic over phrases or their synonyms, defining when and how a change should propagate based on the given change rationale. Based on the identified change propagation conditions, the approach prioritizes requirements by their likelihood of being affected. The assigned score reflects the probability that a given requirement will be impacted by the change associated with the corresponding propagation condition.
Their approach applies text chunking to identify noun phrases (NPs) and verb phrases (VPs) in both the change rationale and the candidate requirement. By detecting phrase-level changes and categorizing them as additions or deletions, the method enables more context-aware impact analysis. For each candidate requirement, it checks whether the changed phrases (NPs/VPs) appear in that requirement under specific change-propagation conditions (e.g., subject/object overlap, semantic similarity). Each satisfied propagation condition contributes to the impact score. The final likelihood score for a candidate requirement depends on the matched propagation conditions defined in the change rationale. The system generates a ranked list of affected requirements for a given change rationale, reducing manual effort by allowing analysts to focus on top-ranked requirements. This approach relies heavily on human-in-the-loop involvement.
\par Alsalemi~\cite{alsalemi2017systematic} conducted a systematic literature review focused on predicting requirements volatility. According to their research, only a few papers have addressed predicting volatility requirements, and most articles have focused on the causes of requirements change and their effects on project performance. Their work underscores the need for more empirical studies to better address the practical aspects of requirements volatility.
Dhamija~\cite{Dhamija2019} presented a systematic study of advancements in change impact analysis techniques. The study found a need for research into hidden dependencies in software requirements that may not be readily apparent. Techniques for identifying such hidden dependencies among software objects, such as specifications, design, and code, need to be proposed.Rusdianto~\cite{rusdianto2024innovative} recently proposed an automated framework for CIA in software projects leveraging LLMs. Their approach integrates the GPT-4 model with Retrieval-Augmented Generation (RAG) to improve reasoning about software artifacts. The method first performs a semantic search across document chunks and ranks them by similarity to the given change request. Subsequently, the LLM analyzes the change using a ranked list of relevant chunks across four key dimensions: technical, functional, business, and stakeholder impact.Rusdianto~\cite{hey2025requirements} proposed an automated LLM-based system for inter-requirement traceability. They further suggested that the same model could be extended for CIA. Their approach first uses RAG to retrieve semantically related documents for a given requirement. Then the LLM determines whether to establish a trace link between the requirement and the retrieved candidates.
\par Existing studies have primarily relied on rule-based or traditional NLP techniques, such as dependency parsing or chunking.
More recent approaches have leveraged LLM capabilities; however, they have primarily focused on traceability and have not assessed the effectiveness of these methods for CIA in scenarios where a change request with a description, rather than a modified requirement, is provided. Therefore, in this paper, we present \textit{ProReFiCIA}, which leverages LLMs while addressing the specific challenges of CIA. Moreover, \textit{ProReFiCIA} requires only a small amount of labeled training data, making it adaptable across domains without extensive data collection or preparation. To the best of our knowledge, this is the first study to comprehensively investigate prompt engineering for this particular problem. Unlike most prior studies, which focus on a single model, our approach evaluates five leading LLMs.

 \section{Conclusion}~\label{sec:conclusion}
This study introduces \textit{ProReFiCIA}, an automated framework for change impact analysis. \textit{ProReFiCIA} leverages prompt engineering with Cache-Augmented Generation (CAG) on large language models and adds two post-processing steps—refinement and filtering—to improve the accuracy of impact predictions. The study also investigates the performance of various LLMs and their robustness to variations in prompt structure and level of detail. Experimental results reveal that GPT-4o and LLaMa models exhibit the highest robustness compared to DeepSeek, Gemini, and Mistral. Moreover, we show that GPT-4o exhibits more robust behavior than LLaMa with respect to prompt detail, maintaining consistent performance across datasets from different domains and varying levels of complexity. Using two datasets, we found that adjusting the prompt's level of detail for the GPT-4o model can improve its performance across datasets. 
Our results show that using the most robust model, GPT-4o, with a prompt that performs well across datasets identifies 85.7\% of impacted requirements while requiring the inspection of only 3.0\% of requirements. Moreover, incorporating domain knowledge into the prompt via RAG improves this performance to 95.8\%, while slightly increasing the cost to 3.4\%.

For future work, we plan to investigate the types and causes of false positives in more detail and mitigate this challenge.
 \section*{Acknowledgement}
This work is supported by NSERC of Canada under the Discovery and CRC programs, grant agreement No. 957254, the Research Ireland grant 13/RC/2094-2, and the Luxembourg National Research Fund under grant numbers C23\allowbreak/IS\allowbreak/17958091\allowbreak/PLAITO and NCER22/IS/16570468/NCER-FT. 
\section*{Data Availability}
We have made the source code used in this paper publicly available in an online annex~\cite{annex}. However, the \projNameNew~data has not been made public due to its proprietary nature.

\balance
\bibliographystyle{ACM-Reference-Format}
\bibliography{paper}

\end{document}